\documentclass[twocolumn,aps,showpacs]{revtex4}
\usepackage[dvips]{graphics,color}
\usepackage{graphicx}
\usepackage{dcolumn}
\usepackage{bm}
\usepackage{epsfig}

\newcommand{\beq}{\begin{equation}}
\newcommand{\eeq}{\end{equation}}
\newcommand{\beqa}{\begin{eqnarray}}
\newcommand{\eeqa}{\end{eqnarray}}
\newcommand{\bd}[1]{ \mbox{\boldmath $#1$}}

\begin{document}
\def\ii{\'\i}

\title{
A Solvable  Model for Many Quark Systems in QCD Hamiltonains }

\author{Tochtli Y\'epez-Mart\ii nez and P. O. Hess}
\affiliation{
Instituto de Ciencias Nucleares, Universidad Nacional Aut\'onoma de M\'exico, \\
Ciudad Universitaria, Circuito Exterior S/N, \\
A.P. 70-543, 04510 M\'exico D.F. Mexico}
\author{A. P. Szczepaniak}
\affiliation{
Department of Physics and Nuclear Theory Center, \\
Indiana University Bloomington, Indiana, 47405-4202, USA}
\author{O. Civitarese}
\affiliation{
Departamento de F\ii sica, Universidad Nacional La Plata \\
C.C.67 (1900), La Plata, Argentina}

\begin{abstract}
{ Motivated by a canonical, QCD Hamiltonian we propose 
 an effective Hamiltonian to represent an arbitrary number of  quarks in hadronic bags.  
   The structure of the effective Hamiltonian is discussed and the BCS-type solutions that 
 may represent constituent quarks are presented.  
     The single particle orbitals are chosen 
   as  3-dimensional harmonic oscillators and we discuss
   a class of exact solutions that can be obtained when a subset of single-particle basis
 states  is restricted to include a certain number of orbital
  excitations. 
 The general problem, which includes all possible orbital
 states, can also be solved by combining analytical and numerical methods. 
  } 
\end{abstract}

\maketitle

\section{Introduction}

One of  the  main interests in hadronic physics is to construct effective, low energy approximations 
 to  QCD and  to find methods enabling to treat the theory in its non-perturbative domain. 
  Over the years a number of non-relativistic or semi-relativistic  
   models describing quarks in hadronic bound states have been proposed, however,  the  full effect of 
   quark-antiquark pairs has never been considered. This is because such effects lead to a  
     many-body problem that can only be treated in some approximation. In hadronic models such 
      approximations are typically driven by phenomenological considerations rather then QCD itself. 
 Here, instead we construct a simple effective quark Hamiltonian guided  by QCD and 
   look for various classes of solutions which may represent hadronic states. 
    
    A possible connection between QCD and an effective quark  model Hamiltonian was presented 
      in  Ref.~\cite{adam1}.  A confinement scenario  and dynamical 
       chiral symmetry breaking were discussed, however, even in the chiraly broken phase the effective Hamiltonian remains  complicated, {\it i.e} it contains interactions among  an infinite number of particles.  
 In \cite{adam-peter} following certain  assumptions, a simple solution for the $SU(2)$ case 
  was presented. In particular quarks were assumed to be confined   
 in single s-wave orbitals in a finite volume and no dynamical contributions
 from gluons was considered. The resulting effective  Hamiltonian allows for 
  analytic solutions with  low energies saturated by color-neutral physical states and colored 
 states shifted to arbitrarily  high energies.  
   That study was followed  by \cite{rmf} were the limitation on the number of quark orbitals was relaxed  
    and more analytic solutions were found.

In this work we extend the method of \cite{rmf} to include all
orbital levels in the quark sector. The color part is now extended
to $SU(3)$, while in the flavor sector  we keep the approximately mass-degenerate 
 $u$ and $d$  quarks. 
As in \cite{adam-peter} dynamical contributions of gluons are not considered, and the
system is confined to a finite volume. Under these approximations we show how 
 QCD can be solved nearly analytically, if the potential
interaction of the QCD Hamiltonian is replaced by a spatially averaged 
 constant interaction. 

The paper is organized as follows: in Section II the model space
and the Hamiltonian are defined, In Section III we study the analytic solutions for two- 
and three-orbital level systems, in the chiral limit. 
 This is followed by the consideration of  an arbitrary number of
levels, and  the quark mass term. We discuss why in that case 
 it is impossible to find simple analytic solutions, unless an orthogonal transformation
and a BCS-type of mean field scheme is adopted. In section IV we
discuss the possible emergence of  meson and baryon spectra that result from the
 lowest  s- and p-quark orbitals. Conclusions are drawn in
Section V.

\section{The Hamiltonian and its  model space } 

The construction of an effective, QCD-inspired Hamiltonian  that approximates its 
low energy spectrum is based on the following assumptions. 
Due to confinement the domain of  fields is expected to be restricted to a finite  volume 
in space, where individual hadrons are located.  This is
achieved by using the 3-dimensional harmonic oscillator for the
orbital states. The confinement  scale is then related to the  width of the
harmonic oscillator wave functions, $\gamma$ 
De-confinement  transition can then be studied as a limit 
when
$\gamma \rightarrow 0$
since in this basis for a quark  to probe large distances would require
   mixing with a large number of excited states. Alternatively one can use a
3-dimensional box, but, as we shall show later on, the oscillator
representation is mathematically easier to handle. Gluons are not taken into
account dynamically, rather they are acconted for by a static
potential and the effects of a confining potential arise from the use
of a confining harmonic oscillator.  Finally the potential is smeared over the hadron scale and effectively replaced by a  constant. 

\begin{figure}[t]
\centerline{\epsfxsize5.5cm\epsffile{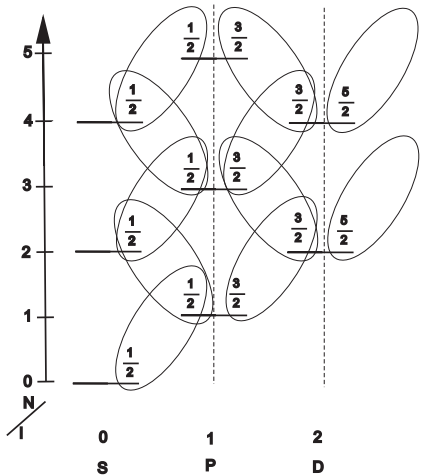}} \caption{
Illustration of the effect of the Hamiltonian on the basis states. The
operators, which appear in the kinetic energy, can be divided into
columns of operators, related to a given spin $j$, which commute
with those of different columns. Only the positive energy states
are plotted. The ellipses indicate which states are connected
through the interaction terms appearing in the kinetic energy.
Columns of different $j$ are separated by a vertical dashed line.
} \label{fig1}
\end{figure}

Most of the restrictions imposed in \cite{adam-peter,rmf} are now
removed. Therein we restricted quarks to be in $SU(2)$ color  but, as we
shall see here, this restriction can be easily eliminated. We
shall also include all the orbital levels, by following the method
developed in \cite{rmf}.

In Fig. \ref{fig1} the model space for the quarks is schematically
presented. Only the
orbital states at positive energy are depicted, while vertical
dashed lines separate columns of equal total spin $j$. As it will
turn out, the Hamiltonian can be written in terms of operators
which act only within each column. This will greatly simplify the
task of finding an analytic procedure.

The fermion creation operator for a quark in the level $\alpha$
($\alpha = \pm\frac{1}{2}$ denotes the upper/lower Dirac level), with angular
momentum $l$ coupled with quark spin-$1/2$ to total spin $j$, is given by

\beqa
{\bd b}^\dagger_{\alpha (N,l\frac{1}{2})j\lambda, cf} & = &
\sum_{m\sigma} (lm,\frac{1}{2}\sigma \mid j \lambda ) {\bd
b}^\dagger_{\alpha Nlm,\sigma cf}.
\label{b-coupl}
\eeqa
The remaining 
indices: N, $\lambda$, $c$ and $f$ refer to the oscillator number,
spin, color and flavor index, respectively. The
$(lm,\frac{1}{2}\sigma \mid j \lambda )$ is an $SU(2)$ Clebsch-Gordan
coefficient. The quarks are in the spin and flavor-spin
representation $\frac{1}{2}$, while for the color part they are in
the triplet irreducible representation. The creation
operator on the right hand side corresponds to a spin-orbit decoupled basis. 
The operator with upper-level indices is defined in the standard way, 
(see Eq.~\ref{b-coupl}), 

\beqa
{\bd b}^{\dagger~\alpha (N,l,\frac{1}{2})j\lambda, cf}
& \equiv &
(-1)^{\frac{1}{2}-\alpha}(-1)^{j-\lambda}
(-1)^{\frac{Y_C}{2}+T_{Cz}}(-1)^{\frac{1}{2}-f} \nonumber \\
&& {\bd b}^\dagger_{-\alpha (N,l,\frac{1}{2})j-\lambda, {\bar
c}-f}.
\eeqa

Here and in the following the  capital
index $C$ denotes globally the color part. $F$ gives the flavor-spin. The 
phase convention is taken from \cite{jutta} (with a corrected sign
in front of $T_z$). The {\it lower-case} $c$ denotes the magnetic
quantum numbers of the color part and is a {\it short hand notation}
for ($Y_C$, $T_C$, $T_{Cz}$), with $Y_C$ being  the color
hypercharge, $T_C$ the color-isospin and $T_{Cz}$ its third
component. The bar over the index $c$ refers to change from 
$c=(Y_C,T_C,T_{Cz})$ to ${\bar c}=(-Y_C,T_C,-T_{Cz})$ (the conjugate index).
In the following we abbreviate $(-1)^ {\frac{Y_C}{2} + T_C}$  as $(-1)^{\chi_c}$.




When the index $\alpha$ has the value  $\alpha = +\frac{1}{2}$
(upper level), the  operators ${\bd b}$ are replaced by ${\bd a}$,
and are referred to as quark operators. In contrast, when $\alpha =
-\frac{1}{2}$ we define them as ${\bd d}$ and as the anti-quark operators. The
${\bd b}$ creation/annihilation operators are then converted into
the ${\bd d}$ annihilation/creation operators. This notation
exploits the Dirac formulation for fermions, where a fermion in an
upper level is denoted as a particle while a fermion hole in
the lower level is denoted as an anti-particle. These and their conjugate will be used to express the effective  Hamiltonian.  The kinetic, Dirac  
  energy is originally  given by

\beqa {\bf
K}=\int{d\bd{x}\bd{\psi}^{\dagger}(\bd{x})[-i{\bd \nabla}
\cdot{\bd \alpha}+\beta m_0] \bd{\psi}(\bd{x})}.
\label{kinetic}
\eeqa
A mass term has been included, in contrast
with \cite{rmf} and the potential  term is given by

\beqa {\bd V} & = & \int{d\bd{x}}d\bd{y}\bd{
\psi}^{\dagger}(\bd{x})T^{a} \bd{ \psi}(\bd{x})
V(|\bd{x}-\bd{y}|)\bd{ \psi}^{\dagger}(\bd{y})T^{a} \bd{
\psi}(\bd{y}). \nonumber \\
\label{pot1} \eeqa 
The static potential $V(|\bd{x}-\bd{y}|)$ simulates the
gluon-quark interaction. The origin of this expression is the
Faddeev-Popov term of the QCD-Hamiltonian \cite{lee}. 
The latter is essentially a color-color, Casimir  interaction and 
 when averaged over the confined hadronic bag 
 it is replaced  by a constant $V_0$,  and becomes  exactly the 
  total color operator \cite{adam-peter}. This can
easily be seen by noting that  $\int{d\bd{x}}\bd{
\psi}^{\dagger}(\bd{x})T^{a} \bd{ \psi}(\bd{x})$ represents the
color charge and a  constant potential separates the integrals over
$\bd{x}$ and $\bd{y}$. 

In what follows, we will first consider the potential interaction,
because it will result in a  very simple spectrum which can later be used to diagonalize the hopping interactions brought by the kinetic energy.  

To do so, we
 expand  the fermion fields ${\bd
 \psi}^{\dagger}$ and ${\bd \psi}$ in this, complete particle basis 
 in terms of the
decoupled fermion creation and annihilation operators, 

\beqa \bd{\psi} ({\bd{x}}) & = & \left(
\begin{array}{c}
\bd{\psi}_1 ({\bd x},\sigma ,c,f) \\
\bd{\psi}_2 ({\bd x},\sigma ,c,f)
\end{array}
\right)
\nonumber \\
{\rm with}
\nonumber \\
{\bd \psi}_{1}({\bd{x}},\sigma ,c,f) & = & \sum_{Nlm}
{\bd b}^{\frac{1}{2}lm,\sigma ,c,f}_{N}R_{Nl}(|\vec{x}|)Y_{lm}(\Omega_x)
\chi_\sigma
\nonumber \\
{\bd \psi}_{2}({\bd{x}},\sigma ,c,f) & = & \sum_{Nlm}
{\bd b}^{-\frac{1}{2}lm,\sigma,c,f}_{N}R_{NL}(|\vec{x}|)Y_{lm}(\Omega_x)
\chi_\sigma. 
\nonumber \\
\label{psi-b}
\eeqa
Here  $\Omega_x$ denotes  the angular components $\theta_x$ and $\phi_x$. of the  postion vector ${\bf x}$.  The fields have components in spin, flavor and color. In  Eqs.~(\ref{kinetic}),(\ref{pot1}) 
 for the kinetic and potential energy,  
 sum over the spin, color and flavor indices is implicit. 

\subsection{The potential}

Substituting into Eq.~(\ref{pot1}) the fermion fields, using Eq.~(\ref{psi-b}) and
transforming  to the coupled fermion creation and annihilation operators  as defined in Eq.~(\ref{b-coupl}) leads to the following form for  the potential interaction (see Appendix A for details)

\beqa
& & {\bd V} = 
\sum_{X} 
\sqrt{(2j_1+1)(2j_3+1)}V(N_i,l_{i},j_{i},L) \nonumber \\
&& (-1)^{M}(-1)^{\frac{Y_C}{2}+T_{Cz}}   \left[\bd{ b}^{\dagger}_{\alpha (N_1,l_{1},\frac{1}{2})j_{1}\lambda_{1},c_{1}f}
\bd{ b}^{\alpha,f}_{(N_2,l_{2},\frac{1}{2})j_{2}\lambda_{2},c_{2}}
\right. \nonumber \\
&& \left.
\langle{j_{1}\lambda_{1},j_{2}\lambda_{2}|LM\rangle}
\langle (1,0)c_{1},(0,1)c_{2}|(1,1)c\rangle_1\right]
\nonumber \\
&& \times \left[\bd{
b}^{\dagger}_{\alpha^{\prime},(N_3,l_{3},\frac{1}{2})j_{3},\lambda_{3},c_{3}
f^{\prime}} \bd{
b}^{\alpha^{\prime},f^{\prime}}_{(N_4,l_{4},\frac{1}{2})j_{4}\lambda_{4},c_{4}}\right.
\nonumber \\
&& \left.
\langle{j_{3}\lambda_{3},j_{4}\lambda_{4}|L-M\rangle}
\langle (1,0)c_{3},(0,1)c_{4}|(1,1) {\bar c}\rangle_1\right]
\nonumber \\
\eeqa 
where $X$ stands for the collection of:  ${N_i,j_{i},\lambda_i,l_i,c_{i},\alpha,f,\alpha^{\prime},f^{\prime},L,M,c}$ quantum numbers. 
The $SU(3)$ Clebsch-Gordan coefficients carry a multiplicity sub-index, $1$,
because the fermion creation and annihilation operators are
coupled to the generators of the color-$SU(3)$. The
coupling corresponds to the  $(1,0)\otimes(0,1)=(0,0)+(1,1)$ decomposition of the product of 
 two  quark representations in which
no multiplicity appears. The convention is taken from
\cite{jutta,draayer1,draayer2} (with a corrected sign in the
$SU(3)$-phase). As before, the indices $\alpha,\alpha'$ are used to label pseudo-spin
components, and $l_i,f,c_i,j_i,\lambda_i$ label orbital angular 
momentum, flavor and color components, total spin and its
projection, respectively. The quantities $V(N_i,l_{i},j_{i},L)$'s
define the intensity of each component of the  interaction in Eq.~\ref{pot1}, and, (before space-averaging) are given by
\begin{widetext}
\beqa
& & V(N_i,l_{i},j_{i},L) =
\frac{1}{(2L+1)}(-1)^{j_{2}+\frac{1}{2}+j_{4}+\frac{1}{2}}
 \frac{3}{2}
\int|\bd{x}|^{2}d|\bd{x}||\bd{y}|^{2}d|\bd{y}|R_{N_1l_{1}}(|\bd{x}|)
R_{N_2l_{2}}(|\bd{x}|)
R_{N_3l_{3}}(|\bd{y}|)R_{N_4l_{4}}(|\bd{y}|)
 \nonumber \\
& & \times  \int^{1}_{-1}d({\rm cos}\theta)P_{L}({\rm cos}\theta)
V(|\bd{x}|,|\bd{y}|,{\rm cos}\theta) \frac{
 \prod^{4}_{i=1}\sqrt{(2l_{i}+1)(2j_{i}+1)}\langle{l_{1}0},l_{2}0|L0\rangle
 \langle{l_{3}0},l_{4}0|L0\rangle
 }{
 \sqrt{(2j_1+1)(2j_3+1)}
 }
\left\{\begin{array}{lll} j_{1}&l_{1}&\frac{1}{2} \\
l_{2}&j_{2}&L\end{array}\right\}
\left\{\begin{array}{lll} j_{3}&l_{3}&\frac{1}{2} \\
l_{4}&j_{4}&L\end{array}\right\}, \nonumber \\
\label{v}
\eeqa
\end{widetext}
The factor $\frac{3}{2}$ originates in the triple-reduced matrix
elements of the color-operator, as explained in Appendix A, and
$\theta$ is the polar angle angle between vectors $\bd{x}$ and $\bd{y}$.

For a space averaged $V(|{\bf x} - {\bf y}| \to V_0$ potential, 
the only  non vanishing contribution comes from $L=0$, which  as a consequence separates
 the integral over ${\bd x}$ and ${\bd y}$ and leads to  $V(N_i,l_{i},j_{i},L)$ to

\begin{equation} 
 V(N_i,l_i,j_i,L=0)  = 
  \frac{V_0}{2}\delta_{L,0}
\delta_{N_1N_2}\delta_{j_1j_2}\delta_{l_1l_2}
\delta_{N_3N_4}\delta_{j_3j_4}\delta_{l_3l_4} \label{pot2}
\end{equation} 

Thus for the potential term in the Hamiltonian we finally have, 

\beqa
V & = & \frac{V_0}{2}
{\bd {\cal C}}_2 (SU(3)), 
\label{s-coupl}
\eeqa
where ${\bd {\cal C}}_2(SU(3)$ is the second order Casimir operator of
$SU(3)$-color, given by

\beqa
{\bd {\cal C}}_2(SU(3)) & = & \frac{3}{2}\sum_{c_1c_2} {\bd C}_{c_1}^{~c_2}
{\bd C}_{c_2}^{~c_1}
\nonumber \\
{\bd C}_{c_1}^{~c_2} & = & \left( {\bd b}^\dagger_{c_1} \cdot {\bd b}^{c_2}
\right) - \frac{\delta_{c_1c_2}}{3} {\bd N}
\nonumber \\
\left( {\bd b}^\dagger_{c_1} \cdot {\bd b}^{c_2} \right) & = &
\sum_{\alpha N l j\lambda f}
{\bd b}^\dagger_{\alpha (Nl,\frac{1}{2})j \lambda c_1 f}
{\bd b}^{\alpha (Nl,\frac{1}{2})j \lambda c_2 f}
\nonumber \\
{\bd N} & = & \sum_{\alpha N l j\lambda c f} {\bd
b}^\dagger_{\alpha (Nl,\frac{1}{2})j \lambda c f} {\bd b}^{\alpha
(Nl,\frac{1}{2})j \lambda c f}.
  \eeqa 
  The ${\bd
C}_{c_1}^{~c_2}$ are the generators of the $SU(3)$-color group.
Its eigenvalue is given by \cite{jutta}, 
\beqa 
\lambda_C^2 +\lambda_C \mu_C + \mu_C^2 + 3\lambda_C + 3\mu_C,
 \eeqa 
 with $(\lambda_C,\mu_C)$ defining the $SU(3)$ irrep. Thus, the potential part simply separates colored from non-colored states. The second-order Casimir operator is equivalent to the color-spin
in $SU(2)$-color discussed in \cite{adam-peter}.

\subsection{The kinetic energy}





We first consider the mass term in Eq.~(\ref{kinetic}). It leads to an operator proportional to
the sum of the quark and antiquark number operators:

\beqa
& m_0 ( {\bd n}_q + {\bd n}_{\bar{q}})
+ m_0\sum_{N l j \lambda c f} 1 &
\nonumber \\
& = m_0\sum_{N j \lambda c f} \left( {\bd
b}^\dagger_{\frac{1}{2} (N, j+\frac{1}{2},\frac{1}{2})j\lambda c
f} {\bd b}^{\frac{1}{2} (N, j+\frac{1}{2},\frac{1}{2})j\lambda c
f} \right. &
\nonumber \\
& \left. -
{\bd b}^\dagger_{-\frac{1}{2} (N, j+\frac{1}{2},\frac{1}{2})j\lambda c f}
{\bd b}^{-\frac{1}{2} (N, j+\frac{1}{2},\frac{1}{2})j\lambda c f}
\right) &
\nonumber \\
& +m_0\sum_{N^{\prime} j \lambda c f} \left( {\bd
b}^\dagger_{\frac{1}{2} (N^{\prime},
j-\frac{1}{2},\frac{1}{2})j\lambda c f} {\bd b}^{\frac{1}{2}
(N^{\prime}, j-\frac{1}{2},\frac{1}{2})j\lambda c f} \right. &
\nonumber \\
& \left. - {\bd b}^\dagger_{-\frac{1}{2} (N^{\prime},
j-\frac{1}{2},\frac{1}{2})j\lambda c f} {\bd b}^{-\frac{1}{2}
(N^{\prime}, j-\frac{1}{2},\frac{1}{2})j\lambda c f} \right). &
\nonumber \\
\label{h0}
\eeqa
The terms with $\alpha = \frac{1}{2}$ count
the number of quarks in the upper level, while the ones with $\alpha =
-\frac{1}{2}$ together with the last term  count the number of holes, thus the
number of anti-quarks, in the lower level. The last term in the
first line is a constant and may be skipped, if convenient. The
${\bd n}_q$ = ${\bd b}^{\dagger}_{\frac{1}{2}\nu} {\bd
b}^{\frac{1}{2}\nu}$ = ${\bd a}^{\dagger}_\nu {\bd a}^\nu$ and
${\bd n}_{\bar q}$ = ${\bd b}^{-\frac{1}{2}\nu} {\bd
b}^\dag_{-\frac{1}{2}\nu}$ =  ${\bd d}^{\dag\nu} {\bd d}_\nu$ are 
the quark and antiquark number operators,  respectively. The 
index $\nu$ is a short-hand notation for  all the indices  which label 
individual creation and annihilation operators.
  

The momentum dependent  part of the kinetic energy, when expressed it in terms of
the fermion creation and annihilation operators, is given by 
(see Appendix  for derivation), 

\beqa
{\bd K} & = & \left( {\widetilde {\bd K}}_+ + {\widetilde
{\bd K}}_- \right). 
\eeqa
where,

\begin{widetext}
\beqa
{\widetilde {\bd K}}_+ & \equiv  &
\sqrt{\gamma} \sum_j \sum_{N=j+\frac{1}{2}}^{\infty} \sum_{\lambda cf}
\left[ \left( \frac{N-j+\frac{3}{2}}{2} \right)^{\frac{1}{2}}
 {\bd
b}^{\dagger}_{\frac{1}{2} (N,j+\frac{1}{2},\frac{1}{2})j\lambda
cf} {\bd b}^{-\frac{1}{2} (N+1,j-\frac{1}{2},\frac{1}{2})j\lambda
cf} \right.
\nonumber \\
&& \left. +\left( \frac{N+j+\frac{3}{2}}{2} \right)^{\frac{1}{2}}
 {\bd
b}^{\dagger}_{\frac{1}{2} (N,j+\frac{1}{2},\frac{1}{2})j\lambda
cf} {\bd b}^{-\frac{1}{2} (N-1,j-\frac{1}{2},\frac{1}{2})j\lambda
cf} + \left( \frac{N-j+\frac{3}{2}}{2} \right)^{\frac{1}{2}} {\bd
b}^{\dagger}_{\frac{1}{2} (N+1,j-\frac{1}{2},\frac{1}{2})j\lambda
cf}  {\bd b}^{-\frac{1}{2}
(N,j+\frac{1}{2},\frac{1}{2})j\lambda cf} \right.
\nonumber \\
&& + \left.
\left( \frac{N+j+\frac{3}{2}}{2} \right)^{\frac{1}{2}}
{\bd b}^{\dagger}_{\frac{1}{2} (N-1,j-\frac{1}{2},\frac{1}{2})j\lambda cf}
 {\bd
b}^{-\frac{1}{2} (N,j+\frac{1}{2},\frac{1}{2})j\lambda cf}
\right]
\nonumber \\
{\widetilde {\bd K}}_- & \equiv  &
\sqrt{\gamma} \sum_j \sum_{N=j+\frac{1}{2}}^{\infty} \sum_{\lambda cf}
\left[ \left( \frac{N-j+\frac{3}{2}}{2} \right)^{\frac{1}{2}}
 {\bd
b}^{\dagger}_{-\frac{1}{2} (N,j+\frac{1}{2},\frac{1}{2})j\lambda
cf} {\bd b}^{\frac{1}{2} (N+1,j-\frac{1}{2},\frac{1}{2})j\lambda
cf} \right.
\nonumber \\
&& \left. \left( \frac{N+j+\frac{3}{2}}{2} \right)^{\frac{1}{2}}
 {\bd
b}^{\dagger}_{-\frac{1}{2} (N,j+\frac{1}{2},\frac{1}{2})j\lambda
cf} {\bd b}^{\frac{1}{2} (N-1,j-\frac{1}{2},\frac{1}{2})j\lambda
cf} + \left( \frac{N-j+\frac{3}{2}}{2} \right)^{\frac{1}{2}} {\bd
b}^{\dagger}_{-\frac{1}{2} (N+1,j-\frac{1}{2},\frac{1}{2})j\lambda
cf}  {\bd b}^{\frac{1}{2}
(N,j+\frac{1}{2},\frac{1}{2})j\lambda cf} \right.
\nonumber \\
&& +
\left. \left( \frac{N+j+\frac{3}{2}}{2} \right)^{\frac{1}{2}}
{\bd b}^{\dagger}_{-\frac{1}{2} (N-1,j-\frac{1}{2},\frac{1}{2})j\lambda cf}
 {\bd
b}^{\frac{1}{2} (N,j+\frac{1}{2},\frac{1}{2})j\lambda cf} \right].
\label{k+-}
\eeqa
\end{widetext}

The ${\widetilde {\bd K}}_+$  operator shifts quarks from the
lower ($\alpha = -1/2$)  to the
 upper  $(\alpha=+1/2$) level, while  ${\widetilde
{\bd K}}_-$ does the opposite. The tilde is used to
distinguish the above operators from actual generators of a $SU(2)$
algebra.

The mass term will be skipped
in most of the cases discussed here for the following reason: the
kinetic energy part without the mass term can be identified as
being proportional to a component of a generator on  an $SU(2)$ algebra.
Thus, it can be diagonalized exactly. The mass term, however, does
not commute with the ${\widetilde {\bd K}}_m$ operators,
which  determine the momentum dependence of kinetic energy and 
destroys exact diagonalizability. 
That they do not commute can also
be seen noting that, for example, the ${\widetilde {\bd K}}_+$
operator creates a particle-hole pair which is equivalent,
as just noted, to raise
the number of quarks and antiquarks. Nevertheless, when we treat
the complete problem, though the $SU(2)$ structure is lost, we can
still diagonalize the kinetic energy part, using the BCS formalism
\cite{ring}.

\section{Analytic and semi-analytic solutions}

In what follows we shall consider two, three, and finally an arbitrary
number of single-quark orbital levels. The reason for choosing this path 
 is to investigate the analytical properties of the
solutions and  if any features in the small basis approximation can be generalized to the  case when any number of quark levels is allowed. 
  In the two- and three-level case, the mass term
is neglected, otherwise no simple solution can be obtained. The
mass term will be included when all orbital levels are taken into
account, since as we show below  in this limit the mass term can be treated together
with the kinetic energy within the BCS formalism.

\subsection{The two level-systems}

This refers to case when, for a given $j$ two orbital levels are considered,  one with $N$
oscillator quanta and with orbital angular momentum
$l=j+\frac{1}{2}$ and the another with either $N^\prime$ = $(N-1)$
or $(N+1)$ quanta and orbital angular momentum $l=j-\frac{1}{2}$.
For example, for $N=1$ and $j=\frac{1}{2}$ this corresponds to the
lowest $p$- plus the lowest or next-to-lowest $s$-orbital.

The momentum dependent part of the kinetic energy can be related to
${\bd K}_\pm$ operators. These operators are
proportional to ${\widetilde {\bd K}}_m$, when restricted to two orbital levels,
and are of the form

\beqa
\bd{K}_+^{jN} & = & \sqrt{2 \cdot 8} \sqrt{2j+1} \nonumber \\
&& \left(
\left[ \bd{b}^\dagger_{\frac{1}{2}(N,l+1,\frac{1}{2})j}
\otimes \bd{b}_{\frac{1}{2}(N^\prime ,l,\frac{1}{2})j}\right]^{000}_{000}
\right.
\nonumber \\
&& \left.
- \left[ \bd{b}^\dagger_{\frac{1}{2}(N^\prime ,l,\frac{1}{2})j}
\otimes \bd{b}_{\frac{1}{2}(N, l+1,\frac{1}{2})j}\right]^{000}_{000}
\right) \nonumber \\
\bd{K}_-^{jN} & = & \sqrt{2 \cdot 8} \sqrt{2j+1} \nonumber \\
&& \left(
-\left[ \bd{b}^\dagger_{-\frac{1}{2}(N, l+1,\frac{1}{2})j}
\otimes \bd{b}_{-\frac{1}{2}(N^\prime ,l,\frac{1}{2})j}\right]^{000}_{000}
\right.
\nonumber \\
&& \left.
+ \left[ \bd{b}^\dagger_{-\frac{1}{2}(N^\prime ,l,\frac{1}{2})j}
\otimes \bd{b}_{-\frac{1}{2}(N, l+1,\frac{1}{2})j}\right]^{000}_{000}
\right), \nonumber \\
\label{k-s} \eeqa The spin-color-flavor coupling denoted by  "$\otimes$" is
defined as

\begin{eqnarray} 
\left[ {\bd A}^{\Gamma_1} \otimes {\bd B}^{\Gamma_2}
\right]^{\Gamma}_{\mu} & = & \sum_{\mu_1\mu_2} \langle \Gamma_1
\mu_1 , \Gamma_2 \mu_2 \mid \Gamma \mu \rangle
A^{\Gamma_1}_{\mu_1} B^{\Gamma_2}_{\mu_2}. \nonumber \\
 \end{eqnarray}  
Here $\Gamma_k$ and $\Gamma$ stand for the quantum numbers denting combined
representation of the spin, color and flavor,
and  $\mu_k$, $\mu$ denote shortly the magnetic quantum numbers,
e.g., $\mu = \lambda c f$.
The expression in 
Eq.~(\ref{k-s}) can be easily derived from Eq.~(\ref{k+-}) by noting
that the coupling to spin-color-flavor singlet implies a
contraction of  all indices. The factor $\sqrt{2 \cdot 8 \cdot
(2j+1)}$, is the square root of the multiplicities of the flavor,
color and spin representation, respectively, and yields the
correct normalization of the singlet representation.

The first index in the creation and annihilation operators refers
to the pseudo-spin component ($\frac{1}{2}$)  for the upper level,
and ($-\frac{1}{2}$) for the lower level. Thus when $l=1$, the kinetic energy 
 couples quarks in
the s-level with quarks in the p-level, of total spin $j=\frac{1}{2}$. 
 When $l=2$, quarks in the p-level,
 with  total spin $j=\frac{3}{2}$, are coupled by the kinetic term to 
quarks in the d-level, also in the  total spin $\frac{3}{2}$.
The orbital values $l$ are given by $j\pm\frac{1}{2}$. The radial
number, $N$, starts from $j+\frac{1}{2}$ and acquires odd (even)
values, while $N^\prime$ has even (odd) values only and satisfies
$N^\prime = N-1$ {\it or} $N+1$. This selection rule is strictly
obeyed only for the harmonic oscillator. Note, that the ${\bd
K}_m$ operators are coupled, for each given spin $j$, to total
color zero. Thus, they commute with the total color-spin operator
of each column $j$.

The ${\bd K}$-operators satisfy the commutation relations

\beqa
& & \left[ \bd{K}^{jN}_+ , \bd{K}^{jN}_-
\right] =
2\bd{K}^{jN}_0 ,
\nonumber \\
& & \left[ \bd{K}^{jN}_0 , \bd{K}^{jN}_\pm
\right] =
\pm \bd{K}^{jN}_\pm ,
\label{comk}
\eeqa
{\it i.e}., for  each combination of $N$ and
$j$  they form  an $SU^{jN}(2)$ pseudo-spin group.
Furthermore  operators with different spins commute.
The operator $K^{jN}_0$ is given by

\begin{equation} 
\bd{K}^{jN}_0  =  \frac{1}{2} \left\{ \left(
\bd{N}^{jN}_{\frac{1}{2}} +
\bd{N}^{jN}_{\frac{1}{2}}
\right) 
- \left(
\bd{N}^{jN}_{-\frac{1}{2}} + \bd{N}^{jN}_{-\frac{1}{2}}
\right) \right\},
\end{equation} 
{\it i.e.} as a  half of the difference between the number of quarks
in the upper  and lower level. The quarks are coupled to total spin $j$.
Note, that the pseudo-spin group $SU^{jN}(2)$
defined by the ${\bd K}_m$ operators is different from the pseudo-spin
group in the original definition, which has been given with reference
to lowering and raising operators within the  same orbital. The
${\bd K}_m$ operators, defined here, raise and lower quarks from one orbital to a different one.

The relation to the operators in Eq.~({\ref{k+-}) is obtained, by  either (a) 
  restricting  to the combination of the orbital $N,
l=j+\frac{1}{2}$ with the orbital $(N-1), l=j-\frac{1}{2}$, or (b) to  the
combination  of the orbital $N, l=j+\frac{1}{2}$ and the
orbital $(N+1), l=j-\frac{1}{2}$.  The relation of the kinetic energy term and the generators of the
$SU(2)$ algebra are is then given by 

for the case (a) 
\begin{equation} 
 {\widetilde {\bd K}}_\pm^{jN}  =  \sqrt{\frac{\gamma
(N-j+\frac{3}{2})}{2}} {\bd K}_\pm^{jN} \equiv 
  A_{aNj}{\bd K}_\pm^{jN},
\label{defab1}
\end{equation}

and for the case (b) 

\begin{equation} 
 {\widetilde {\bd K}}_\pm^{jN}  = \sqrt{\frac{\gamma
(N+j+\frac{3}{2})}{2}} {\bd K}_\pm^{jN} 
\equiv  A_{bNj}{\bd K}_\pm^{jN}.
\label{defab}
\end{equation} 

The operators ${\bd K}_m$  commute with the total color operator in
each column $j$ and, thus, also with the total color operator. This
is because for each given $j$ the  they are  coupled to color zero. As a consequence, the
Hamiltonian can be diagonalized analytically. It is  given by 

\beqa 
{\bd H}^{jN} & = & A_{\kappa Nj}\left( {\bd K}^{jN}_+ + {\bd
K}^{jN}_- \right) + \frac{V_0}{2}{\bd {\cal C}}_2(SU(3)),
\label{ham-2level} 
\eeqa 
with $A_{\kappa Nj}$ ($\kappa$=a,b), 
defined in Eqs.~(\ref{defab1}) and (\ref{defab}). The expression in the parenthesis of the
first term is just ${\bd K}_x^{jN}$, whose eigenvalues are known.
The kinetic part is just then $2{\bf K}_x$. Thus the eigenvalues of
this Hamiltonian are

\beqa
E^{jN} & = &  2A_{\kappa Nj}M_J
\nonumber \\
&& + \frac{V_0}{2} (\lambda_C^2 + \lambda_C\mu_C + \mu_C^2
+ 3\lambda_C + 3\mu_C ),
\label{en-2level}
\eeqa
where $M_J$ is the projection of the pseudo-spin operator, within the
pseudo-spin $J$, onto the $x$-axis and $(\lambda_C , \mu_C )$ define the
irreducible representations (irrep)  the $SU(3)$ color group \cite{draayer1,jutta}. When color
is zero, the irrep is given by $(0,0)$, while for a color octet it is
$(1,1)$, etc. We used the definition of the second order Casimir operator
as given in \cite{draayer1,jutta}. For color less states this reduces to the
simple formula $2A_{\kappa Nj}M_J$.

Fig.~\ref{fig1} illustrates the effect of the kinetic
term discussed above, without
showing the negative energy levels
The negative energy levels are just a copy
of the levels at positive energy inverted to
negative energy of what is shown
in Fig. \ref{fig1}.
Each of the ellipses
represents a combination of orbitals for which the analytic solution given above in Eq.~(\ref{en-2level})  
applies.
The orbitals, which the ellipses connect, are not only the ones at positive
energy, as shown in the figure, but also represent the connection to
the negative energy levels.

\subsection{The three-level system}

The three-level system consists of the following levels:
We
add
to each level, given by $j, N, l=j+\frac{1}{2}$,
two levels with  $l=j-\frac{1}{2}$ and  $(N-1)$ {\it and} $(N+1)$
oscillator quanta. The kinetic energy for this three-level system becomes, 

\begin{equation} 
{ \widetilde {\bd K}} = {\widetilde {\bd K}}_+^{jN} + {\widetilde {\bd K}}_-^{jN}
\end{equation} 
where 
\begin{widetext}
\beqa
{\widetilde {\bd K}}_+^{jN} & = & \sqrt{\gamma} \sum_{\lambda cf}
\left[ \left( \frac{N-j+\frac{3}{2}}{2} \right)^{\frac{1}{2}}
 {\bd
b}^{\dagger}_{\frac{1}{2} (N,j+\frac{1}{2},\frac{1}{2})j\lambda
cf} {\bd b}^{-\frac{1}{2} (N+1,j-\frac{1}{2},\frac{1}{2})j\lambda
cf} \right.
\nonumber \\
&& \left. +\left( \frac{N+j+\frac{3}{2}}{2} \right)^{\frac{1}{2}}
 {\bd
b}^{\dagger}_{\frac{1}{2} (N,j+\frac{1}{2},\frac{1}{2})j\lambda
cf} {\bd b}^{-\frac{1}{2} (N-1,j-\frac{1}{2},\frac{1}{2})j\lambda
cf} + \left( \frac{N-j+\frac{3}{2}}{2} \right)^{\frac{1}{2}} {\bd
b}^{\dagger}_{\frac{1}{2} (N+1,j-\frac{1}{2},\frac{1}{2})j\lambda
cf} {\bd b}^{-\frac{1}{2}
(N,j+\frac{1}{2},\frac{1}{2})j\lambda cf} \right.
\nonumber \\
&& +
\left. \left( \frac{N+j+\frac{3}{2}}{2} \right)^{\frac{1}{2}}
{\bd b}^{\dagger}_{\frac{1}{2} (N-1,j-\frac{1}{2},\frac{1}{2})j\lambda cf}
 {\bd
b}^{-\frac{1}{2} (N,j+\frac{1}{2},\frac{1}{2})j\lambda cf}
\right]
\nonumber \\
{\widetilde {\bd K}}_-^{jN} & = & \sqrt{\gamma} \sum_{\lambda cf}
\left[ \left( \frac{N-j+\frac{3}{2}}{2} \right)^{\frac{1}{2}}
 {\bd
b}^{\dagger}_{-\frac{1}{2} (N,j+\frac{1}{2},\frac{1}{2})j\lambda
cf} {\bd b}^{\frac{1}{2} (N+1,j-\frac{1}{2},\frac{1}{2})j\lambda
cf} \right.
\nonumber \\
&& \left. +\left( \frac{N+j+\frac{3}{2}}{2} \right)^{\frac{1}{2}}
 {\bd
b}^{\dagger}_{-\frac{1}{2} (N,j+\frac{1}{2},\frac{1}{2})j\lambda
cf} {\bd b}^{\frac{1}{2} (N-1,j-\frac{1}{2},\frac{1}{2})j\lambda
cf} + \left( \frac{N-j+\frac{3}{2}}{2} \right)^{\frac{1}{2}} {\bd
b}^{\dagger}_{-\frac{1}{2} (N+1,j-\frac{1}{2},\frac{1}{2})j\lambda
cf} {\bd b}^{\frac{1}{2}
(N,j+\frac{1}{2},\frac{1}{2})j\lambda cf} \right.
\nonumber \\
&& +
\left. \left( \frac{N+j+\frac{3}{2}}{2} \right)^{\frac{1}{2}}
{\bd b}^{\dagger}_{-\frac{1}{2} (N-1,j-\frac{1}{2},\frac{1}{2})j\lambda cf}
 {\bd
b}^{\frac{1}{2} (N,j+\frac{1}{2},\frac{1}{2})j\lambda cf} \right].
\label{k+--3}
\eeqa
\end{widetext}

Note, that this is exactly the same
expression as in Eq.~(\ref{k+-}) but without
summations
over $j$ and $N$.

Defining
\beqa {\bd K}_\pm^{jN} & = & \eta {\widetilde {\bd
K}}_m^{jN}
\nonumber \\
{\bd K}_0^{jN} & = & \eta^{2} {\widetilde {\bd K}}_0^{jN},
\eeqa
it can be verified that, by choosing
the factor
$\eta$
appropriately,  the operators ${\bd K}_m^{jN}$, just like in
the two-level case,
show the standard form of a $SU(2)$ algebra, such that
$\left[ {\bd K}_+^{jN},{\bd K}_-^{jN}\right] =
{\bd K}_0^{jN}$ and 
$\left[ {\bd K}_0^{jN},{\bd K}_\pm^{jN}\right] =
\pm {\bd K}_\pm^{jN}$. 
(see Appendix D for details).
In the three-level case the result then resembles the one
in Eq.~(\ref{ham-2level}) including expression for the energy  given by
Eq.~(\ref{en-2level}). This comes to a complete surprise
because it is not obvious at all that an $SU(2)$ structure is
contained in the relativistic kinetic term in this quasi-particle basis. 

The fact that we found even in such a complicated system an analytic
solution,
is an evidence that probably a complete analytical treatment,
of the whole problem, i.e., including all orbital levels, might be possible,
save a simple numerical solution of a set of equations.
We discuss this case in the following section. 

\subsection{An arbitrary number of levels}

The kinetic energy that is not restricted to operate in a specific subspace of single particle orbitals,   
  is given by Eq.~(\ref{k+-}). The two parts of this kinetic energy, the one
which moves quarks to the upper level and the one which moves
quarks to the lower level, do not satisfy a $SU(2)$ algebra any
more. We will also add a mass term, which by itself destroys any
$SU(2)$ structure once present (see discussion in II.B).
Nevertheless, we can exploit the structure encountered in the two- and three- level case 
  to simplify the problem.

Noting that the total kinetic energy is given by

\beqa
{\bd K} & = & \sum_j
{\bd K}^j
\eeqa
we can select one particular  value $j$ and
solve the problem for it. The complete solution is the sum of the
solutions for all $j$. We introduce a cut-off value ($n$) for $N$,
which can be chosen arbitrarily. Therefore,

\begin{widetext}
\beqa
{\bd K}^{(j,n)} & = & \sum^{n, \Delta N =
2}_{N=j+\frac{1}{2}} \sum^{{\rm min}(n,N+1)}_{N^{\prime}=N-1} \sum_{\lambda cf}
k_{NN^{\prime}}^j
\left({\bd
b}^{\dagger}_{\frac{1}{2}(N,j+\frac{1}{2},\frac{1}{2})j\lambda{cf}}
{\bd
b}^{-\frac{1}{2}(N^{\prime},j-\frac{1}{2},\frac{1}{2})j\lambda{cf}}
+{\bd
b}^{\dagger}_{-\frac{1}{2}(N,j+\frac{1}{2},\frac{1}{2})j\lambda{cf}}
{\bd
b}^{\frac{1}{2}(N^{\prime},j-\frac{1}{2},\frac{1}{2})j\lambda{cf}}\right)
\nonumber \\
&&+\sum^{n,\Delta{N=2}}_{N=j+\frac{1}{2}}~~~
\sum^{{\max}(n,N+1)}_{N^{\prime}=N-1} \sum_{\lambda cf}
k_{NN^{\prime}}^{j*} \left({\bd
b}^{\dagger}_{\frac{1}{2}(N^{\prime},j-\frac{1}{2},\frac{1}{2})j\lambda{cf}}
{\bd b}^{-\frac{1}{2}(N,j+\frac{1}{2},\frac{1}{2})\lambda{cf}}
+{\bd
b}^{\dagger}_{-\frac{1}{2}(N^{\prime},j-\frac{1}{2},\frac{1}{2})j\lambda{cf}}
{\bd
b}^{\frac{1}{2}(N,j+\frac{1}{2},\frac{1}{2})j\lambda{cf}}\right)
\nonumber \\
&& +m_0\sum_{N=j+\frac{1}{2}}^n \sum_{\lambda cf} \left( {\bd b}^\dagger_{\frac{1}{2}
(N, j+\frac{1}{2},\frac{1}{2})j\lambda c f} {\bd b}^{\frac{1}{2}
(N, j+\frac{1}{2},\frac{1}{2})j\lambda c f} -{\bd
b}^\dagger_{-\frac{1}{2} (N, j+\frac{1}{2},\frac{1}{2})j\lambda c
f} {\bd b}^{-\frac{1}{2} (N, j+\frac{1}{2},\frac{1}{2})j\lambda c
f} \right)
\nonumber \\
&& + m_0\sum_{N^{\prime}=j-\frac{1}{2}}^n \sum_{\lambda cf} \left( {\bd
b}^\dagger_{\frac{1}{2} (N^{\prime},
j-\frac{1}{2},\frac{1}{2})j\lambda c f} {\bd b}^{\frac{1}{2}
(N^{\prime}, j-\frac{1}{2},\frac{1}{2})j\lambda c f} -{\bd
b}^\dagger_{-\frac{1}{2} (N^{\prime},
j-\frac{1}{2},\frac{1}{2})j\lambda c f} {\bd b}^{-\frac{1}{2}
(N^{\prime}, j-\frac{1}{2},\frac{1}{2})j\lambda c f} \right).
\label{k+-new}
\eeqa
\end{widetext}
In the last two terms, the step size of $N$ and $N^\prime$ is 2. We
have skipped a trivial constant at the end of Eq.~(\ref{k+-new}), which
will be included  in the BCS formalism that we treat later. The
factors $k_{NN^{\prime}}^j$ can be read  from Eq.~(\ref{k+-}).

The kinetic energy can now be rewritten as

\begin{widetext}
\beqa
{\bd K}^{(j,n)} & = & \sum^{\Delta N =
2, n}_{N=j+\frac{1}{2}} \sum^{{\rm min}(n,N+1)}_{N^{\prime}=N-1}
\sum_{\lambda{cf}} \mid
k_{NN^{\prime}}^j\mid \left( {\bd
b}^{\dagger}_{\frac{1}{2}(N,j+\frac{1}{2},\frac{1}{2})j\lambda{cf}}
{\bd
b}^{-\frac{1}{2}(N^{\prime},j-\frac{1}{2},\frac{1}{2})j\lambda{cf}}
+ {\bd
b}^{\dagger}_{-\frac{1}{2}(N,j+\frac{1}{2},\frac{1}{2})j\lambda{cf}}
{\bd
b}^{\frac{1}{2}(N^{\prime},j-\frac{1}{2},\frac{1}{2})j\lambda{cf}}\right)
\nonumber \\
&&+\sum^{\Delta{N=2},n}_{N=j+\frac{1}{2}}
\sum^{{\rm min}(n,N+1)}_{N^{\prime}=N-1} \sum_{\lambda{cf}}
\mid k_{NN^{\prime}}^j\mid
\left({\bd
b}^{\dagger}_{\frac{1}{2}(N^{\prime},j-\frac{1}{2},\frac{1}{2})j\lambda{cf}}
{\bd
b}^{-\frac{1}{2}(N,j+\frac{1}{2},\frac{1}{2})\lambda{cf}} +{\bd
b}^{\dagger}_{-\frac{1}{2}(N^{\prime},j-\frac{1}{2},\frac{1}{2})j\lambda{cf}}
{\bd
b}^{\frac{1}{2}(N,j+\frac{1}{2},\frac{1}{2})j\lambda{cf}}\right)
\nonumber \\
&&+m_0\sum_{N=j+\frac{1}{2}}^{\Delta N=2,n} \sum_{\lambda{cf}}
\left(
{\bd b}^\dagger_{\frac{1}{2} (N, j+\frac{1}{2},\frac{1}{2})j\lambda c f}
{\bd b}^{\frac{1}{2} (N, j+\frac{1}{2},\frac{1}{2})j\lambda c f}
-{\bd b}^\dagger_{-\frac{1}{2} (N, j+\frac{1}{2},\frac{1}{2})j\lambda c f}
{\bd b}^{-\frac{1}{2} (N, j+\frac{1}{2},\frac{1}{2})j\lambda c f}
\right)
\nonumber \\
&& + m_0\sum_{N^\prime=j-\frac{1}{2}}^{\Delta N^\prime=2,n} \sum_{\lambda{cf}}
\left(
{\bd b}^\dagger_{\frac{1}{2} (N^\prime, j-\frac{1}{2},\frac{1}{2})j\lambda c f}
{\bd b}^{\frac{1}{2} (N^\prime, j-\frac{1}{2},\frac{1}{2})j\lambda c f}
- {\bd b}^\dagger_{-\frac{1}{2} (N^\prime, j-\frac{1}{2},\frac{1}{2})j\lambda c f}
{\bd b}^{-\frac{1}{2} (N^\prime, j-\frac{1}{2},\frac{1}{2})j\lambda c f}
\right). \label{k+-new-2}
\eeqa
\end{widetext}
while the mass term is unaffected.
Next, we apply a  unitary transformation \beqa {\bd
b}^{\dagger}_{\pm\frac{1}{2}(N,j+\frac{1}{2},
\frac{1}{2})j\lambda{cf}} & = & \sum_{k}\alpha_{jNk}^{*}
\widehat{{\bd b}}^{\dagger}_{\pm\frac{1}{2}
(k,j+\frac{1}{2},\frac{1}{2})j\lambda cf}
\nonumber\\
{\bd b}^{\dagger}_{\pm\frac{1}{2}(N^{\prime},j-\frac{1}{2},
\frac{1}{2})j\lambda{cf}} & = & \sum_{q}\beta_{jN^{\prime}q}^{*}
\widehat{{\bd b}}^{\dagger}_{\pm\frac{1}{2}
(q,j-\frac{1}{2},\frac{1}{2})j\lambda cf} {\label{def2}}.
\eeqa
The new operators have to satisfy also the fermion
anti-commutation rules, which imposes a constriction on the
$\alpha$'s and $\beta$'s. In particular for  the operators with orbital spin
$l=j+\frac{1}{2}$ we have

\beqa
& \left\{ {\bd
b}^{\dagger}_{\pm\frac{1}{2}(N_1,j+\frac{1}{2},
\frac{1}{2})j\lambda{c_1f_2}},
{\bd b}^{\pm\frac{1}{2}(N_2,j+\frac{1}{2},\frac{1}{2})j\lambda{c_2f_2}}
\right\} &
\nonumber \\
& = \delta_{N_1N_2}\delta_{c_1c_2}\delta_{f_1f_2} &
\nonumber\\
& \Longrightarrow\sum_{k}\alpha_{jN_1k_1}^*\alpha_{jN_2k_2}  =
\delta_{N_1N_2}\delta_{k_1k_2} &.
{\label{suma alphas =1}}
\eeqa
while for $l=j-\frac{1}{2}$, 

\beqa
& \left\{{\bd b}^{\dagger}_{\pm\frac{1}{2}(N_1^{\prime},j-\frac{1}{2},
\frac{1}{2})j\lambda{cf}},
{\bd b}^{\pm\frac{1}{2}(N_2^{\prime},j-\frac{1}{2},\frac{1}{2})
j\lambda{cf}}\right\} &
\nonumber \\
& = \delta_{N^\prime_1N^\prime_2}\delta_{c_1c_2}\delta_{f_1f_2} &
\nonumber\\
& \Longrightarrow\sum_{q}\beta_{jN_1^{\prime}q_1}^*\beta_{jN_2^{\prime}q_2}
=  \delta_{N_1N_2}\delta_{q_1q_2} &.
{\label{suma betas =1}}
\eeqa

\begin{figure}[!tbh]
\begin{center}
\includegraphics[width=8.5cm, height=6.5cm]{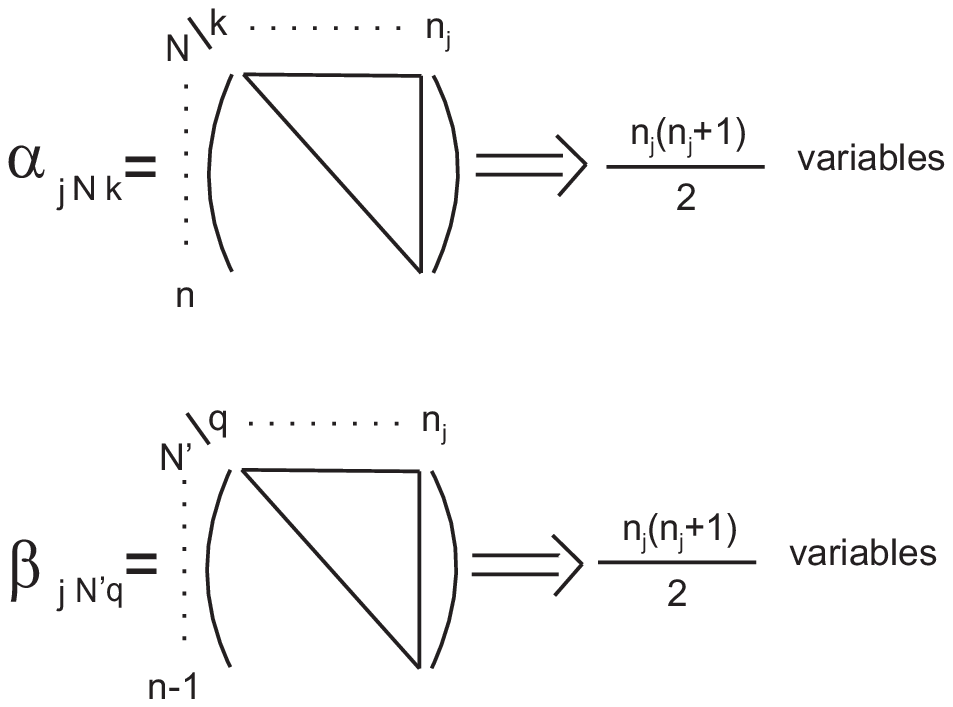}
\end{center}
\caption{Structure of the matrices $\alpha$
y $\beta$. The $n_j$ gives the dimension of the matrix. Though $N$ goes from
$j+\frac{1}{2}$ up to $n$, the maximal number of $N$, considering that $N$
makes steps in two, the number of rows are exactly $n_j$. the relation of
$n_j$ to $n$ is given by
$n_j=\frac{1}{2}\left( n-(j+\frac{1}{2}) \right) +1$.
}
{\label{fig2}}
\end{figure}

A few words are necessary, concerning the variables $\alpha_{jNk}$
and $\beta_{jN^\prime q}$. We cut the space at a given $N=N_{{\rm
max}}=n$ for $j+\frac{1}{2}$, such that the maximal $N^\prime$
equals $N_{{\rm max}}-1$.
Choosing a given integer $n$ as the maximal number for $N$, the range
of $k$ and $q$ is from 1 to $n_j$ (see Fig.~\ref{fig2}), where
$n_j$ is the number of levels for $l=j\pm\frac{1}{2}$. For example,
in the two level system and $j=\frac{1}{2}$, we consider one $p$
orbital (orbital spin $j+\frac{1}{2}$ and $N=1$) and one $s$
orbital (orbital spin $j-\frac{1}{2}$ and $N^\prime =0$). In this
case, $n_j=1$ and $k=q=1$. Using the symmetry of the matrices, we get for
$\alpha_{jNk}$ a total of $n_j(n_j+1)/2$ linear independent elements,
and also for $\beta_{jNq}^{\prime}$. In order to determine all of
them, we need $n_j(n_j+1)$ conditions, as we shall explain later on.
Applying these transformations, the kinetic energy reads

\begin{widetext}
\beqa {\bd K}^j & = & \sum_{kq} \sum_{\lambda cf}
\widetilde{k}^j_{kq}
\left(\widehat{{\bd
b}}^{\dagger}_{\frac{1}{2} (k,j+\frac{1}{2},\frac{1}{2})j\lambda
cf } \widehat{{\bd b}}^{-\frac{1}{2}
(q,j-\frac{1}{2},\frac{1}{2})j\lambda cf}
+\widehat{{\bd b}}^{\dagger}_{-\frac{1}{2}
(k,j+\frac{1}{2},\frac{1}{2})j\lambda cf} \widehat{{\bd
b}}^{\frac{1}{2} (q,j-\frac{1}{2},\frac{1}{2})j\lambda cf}\right)
\nonumber\\
&&+\sum_{kq} \sum_{\lambda cf}
\widetilde{k}_{kq}^{j}
\left(\widehat{{\bd b}}^{\dagger}_{\frac{1}{2}
(q,j-\frac{1}{2},\frac{1}{2})j\lambda cf}
\widehat{{\bd b}}^{-\frac{1}{2}
(k,j+\frac{1}{2},\frac{1}{2})j\lambda cf} +\widehat{{\bd
b}}^{\dagger}_{-\frac{1}{2} (q,j-\frac{1}{2},\frac{1}{2})j\lambda
cf} \widehat{{\bd b}}^{\frac{1}{2}
(k,j+\frac{1}{2},\frac{1}{2})j\lambda cf}\right)
\nonumber \\
&&+\sum_{k} \sum_{\lambda cf}
m_{0,k,j+\frac{1}{2}} \left(
\widehat{{\bd b}}^\dagger_{\frac{1}{2}
(k,j+\frac{1}{2},\frac{1}{2})j\lambda cf}
\widehat{{\bd b}}^{\frac{1}{2}
(k,j+\frac{1}{2},\frac{1}{2})j\lambda cf}
-\widehat{{\bd b}}^\dagger_{-\frac{1}{2}
(k,j+\frac{1}{2},\frac{1}{2})j\lambda cf}
\widehat{{\bd b}}^{-\frac{1}{2}
(k,j+\frac{1}{2},\frac{1}{2})j\lambda cf} \right)
\nonumber \\
&& +\sum_{q} \sum_{\lambda cf}
m_{0,k,j-\frac{1}{2}}\left(
\widehat{{\bd b}}^\dagger_{\frac{1}{2}
(q,j-\frac{1}{2},\frac{1}{2})j\lambda cf} \widehat{{\bd
b}}^{\frac{1}{2} (q,j-\frac{1}{2},\frac{1}{2})j\lambda cf} -
\widehat{{\bd b}}^\dagger_{-\frac{1}{2}
(q,j-\frac{1}{2},\frac{1}{2})j\lambda cf} \widehat{{\bd
b}}^{-\frac{1}{2} (q,j-\frac{1}{2},\frac{1}{2})j\lambda cf}
\right).
 \eeqa
\end{widetext}
with

\beqa
m_{0,k,j-\frac{1}{2}}&=&\sum_{N^\prime=0,2,4,...}^{n-1}m_{0}|\beta_{jN^{\prime}k}|^{2}\nonumber\\
m_{0,k,j+\frac{1}{2}}&=&\sum_{N=1,3,5,..}^{n}m_{0}|\alpha_{jNk}|^{2}.
{\label{m_{0,s,p}}}. 
\eeqa
and the factors ${\widetilde k}^j_{kq}$ are given by
\beqa
\widetilde{k}_{kq}^{j} & = & \sum^{\Delta
N=2,n}_{N=j+\frac{1}{2}} \sum^{{\rm min}(n,N+1)}_{N^{\prime}=N-1}
|k_{NN^{\prime}}^j|\alpha^{*}_{jNk}\beta_{jN^{\prime}q}.
\eeqa
In order that the kinetic energy, for a given $j$ value, is
diagonal in the orbital index, we require
\beqa
{\widetilde k}^j_{kq} & = & 0~ {\rm for}~  k~ \ne q.
\eeqa

These leads to $n_j(n_j-1)$ conditions which together with the $2n_j$
 normalization conditions, $\sum_k {\mid \alpha_{jNk}
\mid}^2=1$ and $\sum_q {\mid \beta_{jN^\prime q} \mid}^2=1$, leads to 
  $n_j(n_j+1)$ equations required to  
 determine all variables $\alpha_{jNk}$ and $\beta_{jN^\prime q}$.

It shows that at the end, when all restrictions are fulfilled, we
arrive at new orbitals labeled by the index $k=1,2,..,n_j$, which
are divided into particles and anti-particles. Note that, when only
the diagonal components of ${\widetilde k}_{kq}^j$ are different
from zero, the kinetic energy contributes to $n_j$ new orbitals for
each orbital angular momentum $l=j+\frac{1}{2}$ and
$l=j-\frac{1}{2}$. By introducing the index $k$, the final form of
the kinetic energy reads

\begin{widetext}
\beqa
{\bd K}^j & = & \sum_{k}\sum_{\lambda cf}\widetilde{k}^j_{kk}
\left(\widehat{{\bd
b}}^{\dagger}_{\frac{1}{2} (k,j+\frac{1}{2},\frac{1}{2})j\lambda
cf } \widehat{{\bd b}}^{-\frac{1}{2}
(k,j-\frac{1}{2},\frac{1}{2})j\lambda cf}
+\widehat{{\bd b}}^{\dagger}_{-\frac{1}{2}
(k,j+\frac{1}{2},\frac{1}{2})j\lambda cf} \widehat{{\bd
b}}^{\frac{1}{2} (k,j-\frac{1}{2},\frac{1}{2})j\lambda cf}\right)
\nonumber\\
&&+\sum_{k}\sum_{\lambda cf}\widetilde{k}_{kk}^{j}
\left(\widehat{{\bd b}}^{\dagger}_{\frac{1}{2}
(k,j-\frac{1}{2},\frac{1}{2})j\lambda cf}
\widehat{{\bd b}}^{-\frac{1}{2}
(k,j+\frac{1}{2},\frac{1}{2})j\lambda cf} +\widehat{{\bd
b}}^{\dagger}_{-\frac{1}{2} (k,j-\frac{1}{2},\frac{1}{2})j\lambda
cf} \widehat{{\bd b}}^{\frac{1}{2}
(k,j+\frac{1}{2},\frac{1}{2})j\lambda cf}\right)
\nonumber \\
&&+\sum_{k}\sum_{\lambda cf}m_{0,k,j+\frac{1}{2}} \left(
\widehat{{\bd b}}^\dagger_{\frac{1}{2}
(k,j+\frac{1}{2},\frac{1}{2})j\lambda cf}
\widehat{{\bd b}}^{\frac{1}{2}
(k,j+\frac{1}{2},\frac{1}{2})j\lambda cf}
-\widehat{{\bd b}}^\dagger_{-\frac{1}{2}
(k,j+\frac{1}{2},\frac{1}{2})j\lambda cf}
\widehat{{\bd b}}^{-\frac{1}{2}
(k,j+\frac{1}{2},\frac{1}{2})j\lambda cf} \right)
\nonumber \\
&& +\sum_{k}\sum_{\lambda cf} m_{0,k,j-\frac{1}{2}}\left(
\widehat{{\bd b}}^\dagger_{\frac{1}{2}
(k,j-\frac{1}{2},\frac{1}{2})j\lambda cf} \widehat{{\bd
b}}^{\frac{1}{2} (k,j-\frac{1}{2},\frac{1}{2})j\lambda cf} -
\widehat{{\bd b}}^\dagger_{-\frac{1}{2}
(k,j-\frac{1}{2},\frac{1}{2})j\lambda cf} \widehat{{\bd
b}}^{-\frac{1}{2} (k,j-\frac{1}{2},\frac{1}{2})j\lambda cf}
\right). \eeqa
\end{widetext}

The number-operators of quarks and anti-quarks ($n_q$ and
$n_{{\bar q}}$) are invariant under the  unitary transformation described above.
 They just transform into the new operators with that also represent 
 number operators of quarks or anti-quarks, the latter
represented by holes in the lower levels. The same is true for the mass term -- it 
 does not change under this transformation.  

Up to this point, however, fermions with the orbital angular momentum $l=j \pm \frac{1}{2}$
are still mixed-in by the kinetic energy.
The final diagonalization will be achieved, applying a BCS transformation
\cite{ring}. The transformation is given by

\beqa
\widehat{{\bd b}}^{\frac{1}{2} (k,
j+\frac{1}{2},\frac{1}{2})j\lambda c f} & = &
c_{j-\frac{1}{2},k}{\bd
b}_{j+\frac{1}{2}}^{(k,j)\lambda{cf}}-s_{j-\frac{1}{2},k} {\bd
d}_{j-\frac{1}{2}}^{\dagger(k,j)\lambda{cf}}
\nonumber\\
\widehat{{\bd b}}^{-\frac{1}{2} (k,
j+\frac{1}{2},\frac{1}{2})j\lambda c f} & = &
s_{j+\frac{1}{2},k}{\bd
b}_{j-\frac{1}{2}}^{(k,j)\lambda{cf}}+c_{j+\frac{1}{2},k}
d_{j+\frac{1}{2}}^{\dagger(k,j)\lambda{cf}}
\nonumber\\
\widehat{{\bd b}}^{\dagger}_{\frac{1}{2} (k,
j+\frac{1}{2},\frac{1}{2})j\lambda c f} & = &
c_{j-\frac{1}{2},k}{\bd
b}^{\dagger}_{j+\frac{1}{2}(k,j)\lambda{cf}}-s_{j-\frac{1}{2},k}
{\bd d}_{j-\frac{1}{2}(k,j)\lambda{cf}}
\nonumber\\
\widehat{{\bd b}}^{\dagger}_{-\frac{1}{2}
(k, j+\frac{1}{2},\frac{1}{2})j\lambda c f} & = &
s_{j+\frac{1}{2},k}{\bd
b}^{\dagger}_{j-\frac{1}{2}(k,j)\lambda{cf}}
+c_{j+\frac{1}{2},k}{\bd d}_{j+\frac{1}{2}(k,j)\lambda{cf}}
\nonumber\\
\widehat{{\bd b}}^{\frac{1}{2} (k,
j-\frac{1}{2},\frac{1}{2})j\lambda c f} & = &
c_{j+\frac{1}{2},k}b_{j-\frac{1}{2}}^{(k,j)\lambda{cf}}-s_{j+\frac{1}{2},k}
{\bd d}_{j+\frac{1}{2}}^{\dagger(k,j)\lambda{cf}}
\nonumber\\
\widehat{{\bd b}}^{\frac{1}{2} (k,
j-\frac{1}{2},\frac{1}{2})j\lambda c f} & = &
s_{j-\frac{1}{2},k}{\bd
b}_{j+\frac{1}{2}}^{(k,j)\lambda{cf}}+c_{j-\frac{1}{2},k} {\bd
d}_{j-\frac{1}{2}}^{\dagger(k,j)\lambda{cf}}
\nonumber\\
\widehat{{\bd
b}}^{\dagger}_{\frac{1}{2}(k,j-\frac{1}{2},\frac{1}{2})j\lambda{cf}}
& = & c_{j+\frac{1}{2},k}{\bd
b}^{\dagger}_{j-\frac{1}{2}(k,j)\lambda{cf}}
-s_{j+\frac{1}{2},k}{\bd d}_{j+\frac{1}{2}(k,j)\lambda{cf}}
\nonumber\\
\widehat{{\bd b}}^{\dagger}_{\frac{1}{2} (k,
j-\frac{1}{2},\frac{1}{2})j\lambda c f} & = &
s_{j-\frac{1}{2},k}{\bd
b}^{\dagger}_{j+\frac{1}{2}(k,j)\lambda{cf}}
+c_{j-\frac{1}{2},k}{\bd d}_{j-\frac{1}{2}(k,j)\lambda{cf}}.
\nonumber\\
{\label{trans
BCS_{k}}}
\eeqa
with $c \equiv \cos(\theta)$ and $s \equiv \sin(\theta)$, and $\theta$ being the Bogolubov  angle. 
Applying this rotation to the  kinetic energy term and including  the mass term,
leads to
(from now on we skip the upper index $j$ in $k_{kk}^j$, because from
the context it is clear that we work in a fixed column denoted by $j$)

\begin{widetext}
\beqa K^{j}_{BCS}&=&\int{d\bd{x}}\bd{\psi}^{\dagger}(\bd{x})
(-i\bd{\alpha}\cdot\bd{\nabla}+\bd{\beta}{m_{0}})\bd{\psi}(\bd{x})
\nonumber\\
&=&\sum_{\lambda{cf}}\sum_{k}
\left\{\left[2\widetilde{k}_{kk}s_{j+\frac{1}{2},k}c_{j+\frac{1}{2},k}
+m_{0,k,j-\frac{1}{2}}c_{j+\frac{1}{2},k}^{2}-m_{0,k,j+\frac{1}{2}}s_{j+\frac{1}{2},k}^{2}\right]
b^{\dagger}_{j-\frac{1}{2}(kj)\lambda{cf}}b_{j-\frac{1}{2}}^{(kj)\lambda{cf}}\right.
\nonumber\\
&+&\left.\left[2\widetilde{k}_{kk}
s_{j-\frac{1}{2},k}c_{j-\frac{1}{2},k}
+m_{0,k,j-\frac{1}{2}}c_{j-\frac{1}{2},k}^{2}-m_{0,k,j+\frac{1}{2}}s_{j-\frac{1}{2},k}^{2}\right]
d^{\dagger(kj)\lambda{cf}}_{j-\frac{1}{2}}d_{j-\frac{1}{2}(kj)\lambda{cf}}\right.
\nonumber\\
&+&\left.\left[2\widetilde{k}_{kk}
s_{j-\frac{1}{2},k}c_{j-\frac{1}{2},k}
+m_{0,k,j+\frac{1}{2}}c_{j-\frac{1}{2},k}^{2}-m_{0,k,j-\frac{1}{2}}s_{j-\frac{1}{2},k}^{2}\right]
b^{\dagger}_{j+\frac{1}{2}(kj)\lambda{cf}}b_{j+\frac{1}{2}}^{(kj)\lambda{cf}}\right.
\nonumber\\
&+&\left.\left[2\widetilde{k}_{kk}
s_{j+\frac{1}{2},k}c_{j+\frac{1}{2},k}
+m_{0,k,j+\frac{1}{2}}c_{j+\frac{1}{2},k}^{2}-m_{0,k,j-\frac{1}{2}}s_{j+\frac{1}{2},k}^{2}\right]
d^{\dagger(kj)\lambda{cf}}_{j+\frac{1}{2}}d_{j+\frac{1}{2}(kj)\lambda{cf}}\right.
\nonumber\\
&-&\left. (2 \cdot 2 \cdot 3)
\left[m_{0,k,j-\frac{1}{2}}(c_{j-\frac{1}{2},k}^{2}-s_{j+\frac{1}{2},k}^{2})
+m_{0,k,j+\frac{1}{2}}(c_{j+\frac{1}{2},k}^{2}-s_{j-\frac{1}{2},k}^{2})\right.\right.
\nonumber\\
&&+\left.\left.2\widetilde{k}_{kk}\left(
s_{j-\frac{1}{2},k}c_{j-\frac{1}{2},k}
+s_{j+\frac{1}{2},k}c_{j+\frac{1}{2},k}\right)\right]\right.
\nonumber\\
&+&\left.\left[\widetilde{k}_{kk}(c_{j+\frac{1}{2},k}^{2}-s_{j+\frac{1}{2},k}^{2})
-(m_{0,k,j-\frac{1}{2}}+m_{0,k,j+\frac{1}{2}})s_{j+\frac{1}{2},k}c_{j+\frac{1}{2},k}\right]\right.
\nonumber\\
&&\left.\left[b^{\dagger}_{j-\frac{1}{2}(kj)\lambda{cf}}d^{\dagger(kj)\lambda{cf}}_{j+\frac{1}{2}}
+d_{j+\frac{1}{2}(kj)\lambda{cf}}b^{(kj)\lambda{cf}}_{j-\frac{1}{2}}\right]\right.
\nonumber\\
&+&\left.\left[\widetilde{k}_{kk}
(c_{j-\frac{1}{2},k}^{2}-s_{j-\frac{1}{2},k}^{2})-(m_{0,k,j-\frac{1}{2}}
+m_{0,k,j+\frac{1}{2}})s_{j-\frac{1}{2},k}c_{j-\frac{1}{2},k}\right]\right.
\nonumber\\
&&\left.
\left[b^{\dagger}_{j+\frac{1}{2}(kj)\lambda{cf}}d^{\dagger(kj)\lambda{cf}}_{j-\frac{1}{2}}
+d_{j-\frac{1}{2}(kj)\lambda{cf}}b^{(kj)\lambda{cf}}_{j+\frac{1}{2}}\right]\right\}
. \label{k-fin1}
 \eeqa
\end{widetext}


The kinetic energy operator is diagonalized, requiring that the
terms quadratic in the annihilation and creation operators vanish, which leads to the gap equation that determines the  Bogolubov angle, 

\begin{widetext} \beqa
\left[\widetilde{k}_{kk}(c_{j-\frac{1}{2},k}^{2}-s_{j-\frac{1}{2},k}^{2})
-(m_{0,k,j-\frac{1}{2}}
+m_{0,k,j+\frac{1}{2}})s_{j-\frac{1}{2},k}c_{j-\frac{1}{2},k}\right]
& = & 0
\nonumber \\
\left[\widetilde{k}_{kk}(c_{j+\frac{1}{2},k}^{2}-s_{j+\frac{1}{2},k}^{2})
-(m_{0,k,j-\frac{1}{2}}+m_{0,k,j+\frac{1}{2}})s_{j+\frac{1}{2},k}c_{j+\frac{1}{2},k}\right]
& = & 0 .
 \eeqa
\end{widetext}

The diagonalized new operator, for a given column $j$, is finally given by,

\begin{widetext}
\beqa
{\bd K}^{j}_{BCS}&=&\int{d\bd{x}}\bd{\psi}^{\dagger}(\bd{x})
(-i\bd{\alpha}\cdot\bd{\nabla}+\bd{\beta}{m_{0}})\bd{\psi}(\bd{x})\nonumber\\
&=&\sum_{\lambda{cf}}\sum_{k}
\left\{\left[2\widetilde{k}_{kk}
s_{j+\frac{1}{2},k}c_{j+\frac{1}{2},k}
+m_{0,k,j-\frac{1}{2}}c_{j+\frac{1}{2},k}^{2}-m_{0,k,j+\frac{1}{2}}s_{j+\frac{1}{2},k}^{2}\right]
b^{\dagger}_{j-\frac{1}{2}(kj)\lambda{cf}}b_{j-\frac{1}{2}}^{(kj)\lambda{cf}}\right.
\nonumber\\
&+&\left.\left[2\widetilde{k}_{kk}
s_{j-\frac{1}{2},k}c_{j-\frac{1}{2},k}
+m_{0,k,j-\frac{1}{2}}c_{j-\frac{1}{2},k}^{2}-m_{0,k,j+\frac{1}{2}}s_{j-\frac{1}{2},k}^{2}\right]
d^{\dagger(kj)\lambda{cf}}_{j-\frac{1}{2}}d_{j-\frac{1}{2}(kj)\lambda{cf}}\right.
\nonumber\\
&+&\left.\left[2\widetilde{k}_{kk}
s_{j-\frac{1}{2},k}c_{j-\frac{1}{2},k}
+m_{0,k,j+\frac{1}{2}}c_{j-\frac{1}{2},k}^{2}-m_{0,k,j-\frac{1}{2}}s_{j-\frac{1}{2},k}^{2}\right]
b^{\dagger}_{j+\frac{1}{2}(kj)\lambda{cf}}b_{j+\frac{1}{2}}^{(kj)\lambda{cf}}\right.
\nonumber\\
&+&\left.\left[2\widetilde{k}_{kk}
s_{j+\frac{1}{2},k}c_{j+\frac{1}{2},k}
+m_{0,k,j+\frac{1}{2}}c_{j+\frac{1}{2},k}^{2}-m_{0,k,j-\frac{1}{2}}s_{j+\frac{1}{2},k}^{2}\right]
d^{\dagger(kj)\lambda{cf}}_{j+\frac{1}{2}}d_{j+\frac{1}{2}(kj)\lambda{cf}}\right.
\nonumber\\
&-&\left. 12 \left[m_{0,k,j-\frac{1}{2}}(c_{j-\frac{1}{2},k}^{2}-s_{j+\frac{1}{2},k}^{2})
+m_{0,k,j+\frac{1}{2}}(c_{j+\frac{1}{2},k}^{2}-s_{j-\frac{1}{2},k}^{2})\right.\right.
\nonumber\\
&&+\left.\left.2\widetilde{k}_{kk}\left(
s_{j-\frac{1}{2},k}c_{j-\frac{1}{2},k}
+s_{j+\frac{1}{2},k}c_{j+\frac{1}{2},k}\right)\right]\right\}
\label{kin-fin2}
\eeqa
\end{widetext}

Introducing a notation ${\bd A}{\bd B}$ to represent the 
$\sum_{\lambda c f} {\bd A}_{\lambda cf} {\bd B}^{\lambda cf}$ it can be 
rewritten as.

\beqa
{\bd K}^j_{BCS} & = & \sum_k \left\{ \epsilon_{bkj+\frac{1}{2}}
{\bd b}_{j+\frac{1}{2},k}^{\dagger}{\bd b}_{j+\frac{1}{2},k}
+
\epsilon_{dkj+\frac{1}{2}}
{\bd d}_{j+\frac{1}{2},k}^{\dagger}{\bd d}_{j+\frac{1}{2},k}
\right.
\nonumber \\
&& \left. +
\epsilon_{bkj-\frac{1}{2}}
{\bd b}_{j-\frac{1}{2},k}^{\dagger}{\bd b}_{j-\frac{1}{2},k}
+
\epsilon_{dkj-\frac{1}{2}}
{\bd d}_{j-\frac{1}{2},k}^{\dagger}{\bd d}_{j-\frac{1}{2},k}
\right\},
\nonumber \\
\label{ham-fin-bcs}
\eeqa

This allows us to determine the spectrum of such a model
including excited states,  which are to be associated with hadrons.
In particular meson states are constructed from
\beqa
& \left[ {\bd b}_{j\pm\frac{1}{2},k}^{\dagger}\otimes
{\bd b}_{j\pm\frac{1}{2},k^\prime}^{\dagger}
\right]^{j(0,0)_C F}_\mu  &
\nonumber \\
& \left[ {\bd b}_{j\pm\frac{1}{2},k}^{\dagger}\otimes
{\bd b}_{j\mp\frac{1}{2},k^\prime}^{\dagger}
\right]^{j(0,0)_C F}_\mu &
\nonumber \\
& \left[ {\bd d}_{j\pm\frac{1}{2},k}^{\dagger}\otimes
{\bd d}_{j\pm\frac{1}{2},k^\prime}^{\dagger}
\right]^{j(0,0)_C F}_\mu
\nonumber \\
& \left[ {\bd d}_{j\pm\frac{1}{2},k}^{\dagger}\otimes
{\bd d}_{j\mp\frac{1}{2},k^\prime}^{\dagger}
\right]^{j(0,0)_C F}_\mu &
\nonumber \\
& \left[ {\bd b}_{j\pm\frac{1}{2},k}^{\dagger}\otimes
{\bd d}_{j\pm\frac{1}{2},k^\prime}^{\dagger}
\right]^{j(0,0)_C F}_\mu
\nonumber \\
& \left[ {\bd b}_{j\pm\frac{1}{2},k}^{\dagger}\otimes {\bd
d}_{j\mp\frac{1}{2},k^\prime}^{\dagger} \right]^{j(0,0)_C F}_\mu &,
\label{44}
\eeqa
where the $j$ refers to the total spin, $(0,0)_C$ indicates the overall color singlet 
and $F$ gives the total meson flavor. The quantity $\mu$ is a
short-hand notation for all magnetic quantum numbers. The colored
states are shifted to larger energies due to the second order
color operator.
The structure of
the meson spectrum is already quite involved.
Examining (\ref{44}) and apply to these states the energy operator of
(\ref{ham-fin-bcs}), we obtain the meson
spectrum. Because in (\ref{ham-fin-bcs}) no explicit dependence on the
coupling to spin and flavor appears but only on the new
orbital indices $k$ and $k^\prime$, states with {\it the same} spin and
flavor are degenerate in energy, fixing $k$ and $k^\prime$.
Only when the orbital indices $k$ and/or $k^\prime$ change, also
the energy changes. For example, we can expect that the lowest states
($k=k^\prime = 1$) with
flavor-spin 1 and 0 multiplets with $j$ = 0 and 1 are degenerate.

For the baryon spectrum,  the states which correspond to three quark
operators (${\bd b}^\dagger$ or ${\bd d}^\dagger$) should be coupled to the
desired flavor and spin. The nucleon is then described by the
three quark operators which correspond to the lowest energy. Also
here,
$F=\frac{1}{2}$, $j=\frac{1}{2}$ multiplet, containing the
nucleons, will be degenerate with the $F=\frac{3}{2}$,
$j=\frac{3}{2}$ multiplet, containing the
$\Lambda$ mesons.

Each quark operator is a complicated function of the bare
quark operators of Eq.~(\ref{psi-b}). Thus, in therms of the bare operators hadrons 
 states contain a sea of quark-antiquark pairs in the background. 
 
As it can be seen, the kinetic energy operator, including the mass
term, can be diagonalized exactly. In the next subsection we apply
the procedure to a four level system and show how the bilinear
equations, involving the $\alpha$ and $\beta$ matrix elements, can
be solved.

\section{A test case: the treatment of four levels}

We consider the four level case, with $j=\frac{1}{2}$, including
the mass term.  We take the first two $p$ ($N=1,3$) and
first two $s$ ($N^\prime$ = 0, 2) orbital levels. The coefficients
${\tilde k}_{kq}^j$ and $\alpha_{jNk}$, $\beta_{jN^\prime q}$ satisfy
the relations

\beqa
\widetilde{k}_{kq}^j=0, k\neq{q}\\
\sum_{q}|\beta_{jN^{\prime}q}|^{2}=1\\
\sum_{k}|\alpha_{jNk}|^{2}=1
\eeqa

where

\beqa
\widetilde{k}_{kq}^j=
\sum_{{N=1,3}}\sum_{N^{\prime}=N-1}^{{\rm min}(3,N+1)} 
|k_{NN^{\prime}}^j|
\alpha_{Nk}^{*}\beta_{N^{\prime}q}
\eeqa

and  the factors $|k_{NN^{\prime}}^j|$ are given by

\beqa
|k_{NN^{\prime}}^j|&=&\sqrt{\gamma}\left(\sqrt{\frac{N-j+\frac{3}{2}}{2}}\delta_{N^{\prime},N+1}\right.\nonumber\\
&+&\left.
\sqrt{\frac{N+j+\frac{3}{2}}{2}}\delta_{N^{\prime},N-1}\right).
\eeqa

This expression shows that $N=1$ connects to $N^{\prime}=0,2$ and
$N=3$ connects to $N^{\prime}=2$, which is due to the use of the
harmonic oscillator basis. The scale $\sqrt{\gamma}$ for this
calculation is taken to be of the order of $1fm^{-1}$. From here on we drop the
upper index $j$ in $k_{NN^{\prime}}^j$. The conditions  that need to be satisfied  are, 

\beqa
&\alpha_{11}^{2}+\alpha_{12}^{2}=1 \nonumber \\
&\alpha_{12}^{2}+\alpha_{32}^{2}=1 \nonumber \\
&\beta_{01}^{2}+\beta_{02}^{2}=1 \nonumber \\
&\beta_{02}^{2}+\beta_{22}^{2}=1 \nonumber \\
&|k_{10}|\alpha_{11}\beta_{02}+|k_{12}|\alpha_{11}\beta_{22}
+|k_{32}|\alpha_{12}\beta_{22}=0 \nonumber \\
&|k_{10}|\alpha_{12}\beta_{01}+|k_{12}|\alpha_{12}\beta_{02}
+|k_{32}|\alpha_{32}\beta_{02}=0,
\eeqa

where $\alpha$'s and $\beta$'s are real. These
equations are solved according to \cite{double}. The numerical
solution of these equations shows the existence of several sets of
solutions ($\alpha_{11}$, $\alpha_{12}$, $\alpha_{32}$,
$\beta_{01}$, $\beta_{02}$, $\beta_{22}$). These solutions produce
repeated values for the coefficients $\widetilde{k}_{kk}$.
Finally, there are three sets of solutions that produce three
different $\widetilde{k}_{kk}$. These sets are described in Table I.

\begin{widetext}
\begin{center}
\begin{table}
\begin{tabular}{|c|c|c|c|c|c|c|c|c|}\hline
Solution & $\alpha_{11}$ & $\alpha_{12}$ & $\alpha_{32}$ &
$\beta_{01}$ & $\beta_{02}$ & $\beta_{22}$ &
$\widetilde{k}_{11}(GeV)$ & $\widetilde{k}_{22}(GeV)$
\\\hline
1 & -0.53452 & 0.84515 & -0.53452 & 0 & 1 & 0 & 0.18092 & 0.23357 \\
2 & 0.57930 & -0.81511 & 0.57930 & -0.10050 & -0.99493 & -0.10050 & 0.14320 & 0.22184 \\
3 & -0.99004 & -0.14072 & 0.99004 & -0.99276 & 0.12010 & -0.99276 & 0.30450 & 0.32383 \\
\hline
\end{tabular}
\caption{
Relevant solutions for $\alpha$'s, $\beta$'s, for the
 the system of four levels. The total spin of the level is $j=\frac{1}{2}$,
i.e. the first two $s$ and $p$ orbitals are considered. }
\label{tab1}
\end{table}
\end{center}
\end{widetext}

In the transformed basis  the kinetic energy including the mass term can be written as

\vskip 0.5cm

\begin{widetext}
\beqa
{\bd K}^{\frac{1}{2}}_{BCS}&=&\sum_{\lambda{cf}}\sum_{k}
\left\{\left[2\widetilde{k}_{kk}s_{P,k}c_{P,k}
+m_{0,k,S}c_{P,k}^{2}-m_{0,k,P}s_{P,k}^{2}\right]
b^{\dagger}_{S(kj)\lambda{cf}}b_{S}^{(kj)\lambda{cf}}\right.\nonumber\\
&+&\left.\left[2\widetilde{k}_{kk}s_{S,k}c_{S,k}
+m_{0,k,S}c_{S,k}^{2}-m_{0,k,P}s_{S,k}^{2}\right]
d^{\dagger(kj)\lambda{cf}}_{S}d_{S(kj)\lambda{cf}}\right.\nonumber\\
&+&\left.\left[2\widetilde{k}_{kk}s_{S,k}c_{S,k}
+m_{0,k,P}c_{S,k}^{2}-m_{0,k,S}s_{S,k}^{2}\right]
b^{\dagger}_{P(kj)\lambda{cf}}b_{P}^{(kj)\lambda{cf}}\right.\nonumber\\
&+&\left.\left[2\widetilde{k}_{kk}s_{P,k}c_{P,k}
+m_{0,k,P}c_{P,k}^{2}-m_{0,k,S}s_{P,k}^{2}\right]
d^{\dagger(kj)\lambda{cf}}_{P}d_{P(kj)\lambda{cf}}\right.
\nonumber\\
&-&\left. 12 \left[m_{0,k,S}(c_{S,k}^{2}-s_{P,k}^{2})
+m_{0,k,P}(c_{P,k}^{2}-s_{S,k}^{2})
+
2\widetilde{k}_{kk}\left(s_{S,k}c_{S,k}
+s_{P,k}c_{P,k}\right)\right]\right.
\nonumber\\
&+&\left.\left[\widetilde{k}_{kk}(c_{P,k}^{2}-s_{P,k}^{2})
-(m_{0,k,S}+m_{0,k,P})s_{P,k}c_{P,k}\right]
\left[b^{\dagger}_{S(kj)\lambda{cf}}d^{\dagger(kj)\lambda{cf}}_{P}
+d_{P(kj)\lambda{cf}}b^{(kj)\lambda{cf}}_{S}\right]\right.
\nonumber\\
&+&\left.\left[\widetilde{k}_{kk}(c_{S,k}^{2}-s_{S,k}^{2})-(m_{0,k,S}
+m_{0,k,P})s_{S,k}c_{S,k}\right]
\left[b^{\dagger}_{P(kj)\lambda{cf}}d^{\dagger(kj)\lambda{cf}}_{S}
+d_{S(kj)\lambda{cf}}b^{(kj)\lambda{cf}}_{P}\right]\right\}\label{k-fin3}
\eeqa
\end{widetext}

Following the procedure described in Section III.C, we have to
solve the gap equations  for the mixing angles $\theta_{S,k}$ and
$\theta_{P,k}$

\beqa
\left[\widetilde{k}_{kk}(c_{S,k}^{2}-s_{S,k}^{2})-(m_{0,k,S}
+m_{0,k,P})s_{S,k}c_{S,k}\right]=0 \nonumber\\
\left[\widetilde{k}_{kk}(c_{P,k}^{2}-s_{P,k}^{2})-(m_{0,k,S}
+m_{0,k,P})s_{P,k}c_{P,k}\right]=0, \nonumber\\
\eeqa
respectively  and we find immediately that
these equations
imply $\theta_{S,k}=\theta_{P,k}=\theta_{k}$.
The $\theta_k$ values are listed in Table \ref{tab2} for different possible
solutions and are in general not equal.
The masses are given by $m_{0,k,S}=\sum_{N^{\prime}}m_{0}|\beta_{N^{\prime}k}|^{2}$,
$m_{0,k,P}=\sum_{N}m_{0}|\alpha_{Nk}|^{2}$. The bare mass 
$m_{0}=0.008GeV$, is chosen at the  intermediate value between the mass of the
up and down quarks \cite{grand}. Finally, with the values
$\widetilde{k}_{11}$ and $\widetilde{k}_{22}$
obtained (see Table \ref{tab1})
we can calculate the
pair-energies $\epsilon_{S,kk}$, $\epsilon_{P,kk}$ and the ground
state energy $\epsilon_{0,kk}$:

\begin{widetext}
\beqa
&\left[2\widetilde{k}_{kk}s_{P,k}c_{P,k}
+m_{0,S}c_{P,k}^{2}-m_{0,P}s_{P,k}^{2}\right]=\epsilon_{S,kk}\nonumber\\
&\left[2\widetilde{k}_{kk}s_{S,k}c_{S,k}
+m_{0,P}c_{S,k}^{2}-m_{0,S}s_{S,k}^{2}\right]=\epsilon_{P,kk}\nonumber\\
& 12\left[m_{0,S}(c_{S,k}^{2}-s_{P,k}^{2})
+m_{0,P}(c_{P,k}^{2}-s_{S,k}^{2})
+2\widetilde{k}_{kk}\left(s_{S,k}c_{S,k}
+s_{P,k}c_{P,k}\right)\right]=\epsilon_{0,kk},
\eeqa
\end{widetext}

the results are shown in the Table \ref{tab2}:

\begin{widetext}
\begin{center}
\begin{table}
\begin{tabular}{|c|c|c|c|c|c|c|c|c|}\hline
  Sol & $\theta_{1}$ & $\theta_{2}$& $\epsilon_{S,11}$ & $\epsilon_{P,11}$ & $\epsilon_{0,11}$ & $\epsilon_{S,22}$ & $\epsilon_{P,22}$ & $\epsilon_{0,22}$\\\hline
  1 & 0.76331 & 0.76828 & 0.18111 & 0.18111 & 4.34658 & 0.23372 & 0.23372 & 5.60922 \\
  2 & 0.75749 & 0.76737 & 0.14343 & 0.14343 & 3.44223 & 0.22199 & 0.22199 & 5.32784 \\
  3 & 0.77226 & 0.77305 & 0.30461 & 0.30461 & 7.31072 & 0.32394 & 0.32394 & 7.77446 \\ \hline
\end{tabular}
\caption{
Solutions of the BCS equation, using the $\alpha$ and $\beta$ values from Table I. 
 The lowest energy solution is comes from the Set  2, as can be seen by
the $\epsilon_{S,11}$ and $\epsilon_{P,11}$ values.
}
\label{tab2}
\end{table}
\end{center}
\end{widetext}

Taking the solution which yields the lowest energies 
we obtain for the kinetic energy of the Hamiltonian

\beqa
{\bd K}^{\frac{1}{2}}_{BCS} & = &
\epsilon_{S,11} \left( {\bd b}^\dagger_{S(1\frac{1}{2})} {\bd b}^{S(1\frac{1}{2})}
+ {\bd d}^\dagger_{S(1\frac{1}{2})} {\bd d}^{S(1\frac{1}{2})} \right)
\nonumber \\
&& + \epsilon_{P,11} \left( {\bd b}^\dagger_{P(1\frac{1}{2})} {\bd
b}^{P(1\frac{1}{2})} + {\bd d}^\dagger_{P(1\frac{1}{2})} {\bd
d}^{P(1\frac{1}{2})} \right).
 \eeqa 

The energy of the lowest $s$ and $p$ orbital are degenerate. This
is due to the structure of the kinetic energy. A residual
interaction in the potential term should remove it. Single-particle
states belong to the spin $\frac{1}{2}$, color $(1,0)$ and
flavor-spin $\frac{1}{2}$ representations. The antiparticles
belong to the conjugate representations. In order to build the
 low lying  meson states, we have to put one particle and one
anti-particle in the lowest level, with energy $0.143 \mbox{ GeV}$. 
Thus the energy of the lightest  mesons would be expected around $0.286\mbox{  GeV}$, in good agreement with the scale of excitations. Considering that a baryon needs three quarks, in
order to be in a colorless state, one obtains the first states
about $0.429\mbox{ GeV}$, which is by about a factor two lower than the
empirical value.

As shown in section II.A, the potential term is proportional to
the second order Casimir operator of $SU(3)$ color. This term commutes
with the kinetic energy and, thus, one can diagonalize the kinetic
energy independently to the potential term.


\section{Conclusions}

In this work we have analyzed a QCD-inspired Hamiltonian,
considering a constant interaction between quarks. The constant
interaction represents an average over all residual interactions.
All orbital levels were taken into account in the calculations.
The color part was described by a $SU(3)$ group and the flavor
part was described by a $SU(2)$ group (only up and down quarks
have been included in the model). Nevertheless, it can be easily
extended to $SU(3)$, by adding an additional mass-term for strange
quarks to the kinetic energy. No gluons have been included
dynamically. The system was confined to a finite volume, with the
length of an average harmonic oscillator.

The potential interaction resulted in the color-spin operator
(second order Casimir operator of the color-$SU(3)$ group). This
interaction divides the colorless states from the colored ones,
shifting the colored states towards higher energies, being the shift
proportional to $V_0$, the strength of the constant interaction.

The kinetic energy operator could be written in terms of the sum
of the ${\widetilde {\bd K}}_\pm^{jN}$ operators. restricting to two-
and three-level sub-systems and excluding the mass term. We proved
that they satisfy an $SU(2)$ algebra, by adding a ${\bd K}_0^{jN}$
operator which counts the difference between quarks in the upper
and lower orbital levels. This suggested that an analytic or
quasi-analytic solution may exists also for the general case. This
solution was found after introducing an unitary transformation, to
the quark creation and annihilation operators, followed by a
simple BCS-type of transformation.

The main features of the calculated spectrum are:

i)Its structure is already quite rich,  and resembles the physical
meson and baryon spectrum,

ii)the mesons and baryons consist of two and three partons,
respectively, which once transformed into the original quark
operators correspond to a back-ground sea of many quark-antiquarks
pairs.

As shown, though QCD is highly non-perturbative, there is still a
lot of room for finding analytic solutions for specific sectors of
the theory, particularly, by applying algebraic and
non-perturbative transformations based on many-body techniques.

\vskip 1cm
\section*{Acknowledgements}

We acknowledge financial help from DGAPA and from the National
Research Council of Mexico (CONACyT), and the CONICET (Argentina), and the US. Department of Energy.

\section*{Appendix A: Constructing the potential interaction:}

The potential term is given by the color-color interaction

\beqa
{\bd V} & = & \int{d\bd{x}}d\bd{y}\bd{
\psi}^{\dagger}(\bd{x})T^{a} \bd{ \psi}(\bd{x})
V(|\bd{x}-\bd{y}|)\bd{ \psi}^{\dagger}(\bd{y})T^{a}
\bd{ \psi}(\bd{y}), \nonumber \\
\eeqa
where the fermion operators $\bd{\psi}^{\dagger}$ and
$\bd{\psi}$ have the following structure

\begin{widetext}
\beqa
\bd{\psi}^{\dagger}(\bd{x})&=&(\bd{\psi}_{1}^{\dagger}(\bd{x}),\psi_{2}^{\dagger}(\bd{x}))\nonumber\\
&=&\left(\sum_{Nlm\sigma cf} b^{\dagger}_{\frac{1}{2}Nlm\sigma c
f}
R_{Nl}^{*}(\bd{x})Y_{lm}^{*}(\Omega_{x})\chi^{\dagger}_{\sigma},
\sum_{Nlm\sigma cf} b^{\dagger}_{-\frac{1}{2}Nlm\sigma c f}
R_{Nl}^{*}(\bd{x})Y_{lm}^{*}(\Omega_{x})\chi^{\dagger}_{\sigma}\right)\nonumber\\
\bd{\psi}(\bd{x})&=&
\left(\begin{array}{l}
\sum_{Nlm\sigma cf}
b^{\frac{1}{2}Nlm\sigma c
f} R_{Nl}(\bd{x})Y_{lm}(\Omega_{x})\chi_{\sigma} \\
\sum_{Nlm\sigma cf} b^{-\frac{1}{2}Nlm\sigma c f}
R_{Nl}(\bd{x})Y_{lm}(\Omega_{x})\chi_{\sigma}
\end{array}
\right).
\eeqa
\end{widetext}
So the potential term becomes

\begin{widetext}
\beqa
\bd{V}&=&\sum_{\alpha_{i}N_{i},l_{i}m_{i}j_{i}\lambda_{i}\sigma_{i}f_{i}c_{i}LM}
\int d\bd{x}d\bd{y}V(|\bd{x}-\bd{y}|)\nonumber\\
&&\left\{\left[b^{\dagger}_{\alpha_{1}(N_{1},l_{1},\frac{1}{2})j_{1}\lambda_{1}c_{1}f_{1}}
\langle l_{1}m_{1},\frac{1}{2}\sigma_{1}|j_{1}\lambda_{1}\rangle
R_{N_{1}l_{1}}^{*}(\bd{x})Y_{l_{1}m_{1}}^{*}(\Omega_{x})\chi^{\dagger}_{\sigma_{1}}
(T^{a})^{c_{1}}_{c_{2}}\right.\right.\nonumber\\
&&\left.\left.
b^{\alpha_{2}(N_{2},l_{2},\frac{1}{2})j_{2}\lambda_{2}c_{2}f_{2}}
\langle l_{2}m_{2},\frac{1}{2}\sigma_{2}|j_{2}\lambda_{2}\rangle
R_{N_{2}l_{2}}(\bd{x})Y_{l_{2}m_{2}}(\Omega_{x})\chi_{\sigma_{2}}
\delta_{f_{1}f_{2}}\delta_{\alpha_{1}\alpha_{1}}\delta_{\sigma_{1}\sigma_{2}}\right]\right.\nonumber\\
&&\left.\times
\left[b^{\dagger}_{\alpha_{3}(N_{3},l_{3},\frac{1}{2})j_{3}\lambda_{3}c_{3}f_{3}}
\langle l_{3}m_{3},\frac{1}{2}\sigma_{3}|j_{3}\lambda_{3}\rangle
R_{N_{3}l_{3}}^{*}(\bd{y})Y_{l_{3}m_{3}}^{*}(\Omega_{y})\chi^{\dagger}_{\sigma_{3}}
(T^{a})^{c_{3}}_{c_{4}}\right.\right.\nonumber\\
&&\left.\left.
b^{\alpha_{4}(N_{4},l_{4},\frac{1}{2})j_{4}\lambda_{4}c_{4}f_{4}}
\langle l_{4}m_{4},\frac{1}{2}\sigma_{4}|j_{4}\lambda_{4}\rangle
R_{N_{4}l_{4}}(\bd{y})Y_{l_{4}m_{4}}(\Omega_{y})\chi_{\sigma_{4}}
\delta_{f_{3}f_{4}}\delta_{\alpha_{3}\alpha_{4}}\delta_{\sigma_{3}\sigma_{4}}\right]\right\}.
\eeqa
\end{widetext}

The angular part of the double integral is given by the following
expression:

\begin{widetext}
\beqa
\int\int{d\Omega_{x}}{d\Omega_{y}}Y_{l_{1}m_{1}}^{*}(\Omega_{x})Y_{l_{2}m_{2}}(\Omega_{x})
V(|\bd{x}-\bd{y}|)
Y_{l_{3}m_{3}}^{*}(\Omega_{y})Y_{l_{4}m_{4}}(\Omega_{4})
\eeqa
\end{widetext}
where the residual interaction can be written as

\beqa
V(|\bd{x}-\bd{y}|)&=&\sum_{L}A_{L}P_{L}(cos\theta)\nonumber\\
\Rightarrow
A_{L}&=&\left(\frac{2L+1}{2}\right)\int^{1}_{-1}d(cos\theta)P_{L}(cos\theta)V(|\bd{x}-\bd{y}|)\nonumber\\
P_{L}(cos\theta)&=&
\left(\frac{4\pi}{2L+1}\right)\sum_{M}Y_{LM}^{*}(\Omega_{x})Y_{LM}(\Omega_{y}).
\eeqa

The second expression comes from the orthogonality of the Legendre
polynomials and the third is a useful relation for the Legendre
Polynomials. Therefore, the double angular integral can be separated in
two single angular integrals:

\begin{widetext}
\beqa
&\rightarrow&\int\int{d\Omega_{x}}{d\Omega_{y}}Y_{l_{1}m_{1}}^{*}(\Omega_{x})Y_{l_{2}m_{2}}(\Omega_{x})
V(|\bd{x}-\bd{y}|)
Y_{l_{3}m_{3}}^{*}(\Omega_{y})Y_{l_{4}m_{4}}(\Omega_{y})\nonumber\\
&=&\sum_{LM}A_{L}\left(\frac{4\pi}{2L+1}\right)(-1)^{m_{1}+M+m_{3}}
\left[\int{d\Omega_{x}}Y_{l_{1}-m_{1}(\Omega_{x})}Y_{l_{2}m_{2}(\Omega_{x})}Y_{L-M}(\Omega_{x})\right]
\left[\int{d\Omega_{y}}Y_{l_{3}-m_{3}(\Omega_{y})}
Y_{l_{4}m_{4}(\Omega_{y})}Y_{LM}(\Omega_{y})\right]\nonumber\\
&=&\sum_{LM}A_{L}\left(\frac{4\pi}{2L+1}\right)(-1)^{m_{1}+M+m_{3}}
\left[\frac{(2l_{1}+1)(2l_{2}+1)(2L+1)}{4\pi}\right]^{\frac{1}{2}}
\left[\frac{(2l_{3}+1)(2l_{4}+1)(2L+1)}{4\pi}\right]^{\frac{1}{2}}\nonumber\\
&&\times\left(\begin{array}{lll} l_{1} & l_{2} & L\\
0 & 0 & 0\end{array}\right)
\left(\begin{array}{lll} l_{1} & l_{2} & L\\
-m_{1} & m_{2} & -M\end{array}\right)
\left(\begin{array}{lll} l_{3} & l_{4} & L\\
0 & 0 & 0\end{array}\right)
\left(\begin{array}{lll} l_{3} & l_{4} & L\\
-m_{3} & m_{4} & M\end{array}\right).
\eeqa
\end{widetext}

With the above expressions
, the potential term takes the following
form

\begin{widetext}
\beqa
\bd{V}&=&\sum_{\alpha_{i}N_{i},l_{i}m_{i}j_{i}\lambda_{i}\sigma_{i}f_{i}c_{i}LM}
\left[b^{\dagger}_{\alpha_{1}(N_{1},l_{1},\frac{1}{2})j_{1}\lambda_{1}c_{1}f_{1}}
(T^{a})^{c_{1}}_{c_{2}}
b^{\alpha_{2}(N_{2},l_{2},\frac{1}{2})j_{2}\lambda_{2}c_{2}f_{2}}
\chi^{\dagger}_{\sigma_{1}}\chi_{\sigma_{2}}
\delta_{f_{1}f_{2}}\delta_{\alpha_{1}\alpha_{2}}
\delta_{\sigma_{1}\sigma_{2}}\right]
\nonumber\\
&&\left[b^{\dagger}_{\alpha_{3}(N_{3},l_{3},\frac{1}{2})j_{3}\lambda_{3}c_{3}f_{3}}
(T^{a})^{c_{3}}_{c_{4}}
b^{\alpha_{4}(N_{4},l_{4},\frac{1}{2})j_{4}\lambda_{4}c_{4}f_{4}}
\chi^{\dagger}_{\sigma_{3}}\chi_{\sigma_{4}}
\delta_{f_{3}f_{4}}\delta_{\alpha_{3}\alpha_{4}}
\delta_{\sigma_{3}\sigma_{4}}\right]
\nonumber\\
&&
\int |\bd{x}|^{2}d|\bd{x}||\bd{y}|^{2}d|\bd{y}|
R_{N_{1}l_{1}}^{*}(|\bd{x}|) R_{N_{2}l_{2}}(|\bd{x}|)
R_{N_{3}l_{3}}^{*}(|\bd{y}|) R_{N_{4}l_{4}}(|\bd{y}|)
\nonumber\\
&&\int^{1}_{-1}d(cos\theta)P_{L}(cos\theta)V(|\bd{x}-\bd{y}|)
\left(\frac{2L+1}{2}\right)\prod_{i=1}^{4}\sqrt{(2l_{i}+1)(2j_{i}+1)}
\frac{1}{\sqrt{2L+1}}\frac{1}{\sqrt{2L+1}}\nonumber\\
&&(-1)^{m_{1}+M+m_{3}+l_{1}-l_{2}+l_{3}-l_{4}+l_{1}-\frac{1}{2}
-\lambda_{1}+l_{2}-\frac{1}{2}-\lambda_{2}+l_{3}-\frac{1}{2}-\lambda_{3}+l_{4}-\frac{1}{2}-\lambda_{4}}
\langle{l_{1}0},l_{2}0|L0\rangle \langle
l_{3}0,l_{4}0|L0\rangle\nonumber\\
&&\left(\begin{array}{lll} l_{1} & \frac{1}{2} & j_{1}\\
m_{1} & \sigma_{1} & -\lambda_{1}\end{array}\right)
\left(\begin{array}{lll} l_{2} & \frac{1}{2} & j_{2}\\
m_{2} & \sigma_{2} & -\lambda_{2}\end{array}\right)
\left(\begin{array}{lll} l_{1} & l_{2} & L\\
-m_{1} & m_{2} & -M\end{array}\right)\nonumber\\
&&\left(\begin{array}{lll} l_{3} & \frac{1}{2} & j{3}\\
m_{3} & \sigma_{3} & -\lambda_{3}\end{array}\right)
\left(\begin{array}{lll} l_{4} & \frac{1}{2} & j_{4}\\
m_{4} & \sigma_{4} & -\lambda_{4}\end{array}\right)
\left(\begin{array}{lll} l_{3} & l_{4} & L\\
-m_{3} & m_{4} & M\end{array}\right){\label{V with six 3-jcoef}}
\eeqa
\end{widetext}

\vskip 0.5cm

\begin{widetext}
\beqa
&\rightarrow &\sum_{m_{1}\sigma_{1}m_{2}}
(-1)^{-\sigma_{1}-\frac{1}{2}-\sigma_{3}-\frac{1}{2}+M-\lambda_{2}-\frac{1}{2}-\lambda_{4}-\frac{1}{2}}\nonumber\\
&&\left(
\begin{array}{lll}
l_{1} & \frac{1}{2} & j_{1}\\
m_{1} & \sigma_{1} & -\lambda_{1}
\end{array}
\right)
\left(
\begin{array}{lll}
l_{2} & \frac{1}{2} & j_{2}\\
m_{2} & \sigma_{1} & -\lambda_{2}
\end{array}
\right)
\left(
\begin{array}{lll}
l_{1} & l_{2} & L\\
-m_{1} & m_{2} & -M
\end{array}
\right)
\nonumber\\
&=&\sum_{m_{1}\sigma_{1}m_{2}}
\left(
\begin{array}{lll}
j_{1} & l_{1} & \frac{1}{2}\\
-\lambda_{1} & m_{1} & \sigma_{1}
\end{array}
\right)
\left(
\begin{array}{lll}
l_{2} & j_{2}& \frac{1}{2} \\
-m_{2} & \lambda_{2} & -\sigma_{1}
\end{array}
\right)
\left(
\begin{array}{lll}
l_{2} & l_{1} & L\\
m_{2} & -m_{1} &-M
\end{array}
\right)
(-1)^{l_{2}+l_{1}+\frac{1}{2}+m_{2}+m_{1}-\sigma_{1}}
\nonumber\\
&&(-1)^{-\sigma_{1}-\frac{1}{2}-\sigma_{3}-\frac{1}{2}+M-\lambda_{2}-\frac{1}{2}-\lambda_{4}-\frac{1}{2}
-l_{1}-l_{2}-\frac{1}{2}-m_{2}-m_{1}+\sigma_{1}+l_{2}+l_{1}+L}
\nonumber\\
&=&\left(
\begin{array}{lll}
j_{1} & j_{2}& L \\
-\lambda_{1} & \lambda_{2} & -M
\end{array}
\right)
\left\{
\begin{array}{lll}
j_{1} & j_{2}& L \\
l_{2} & l_{1} & \frac{1}{2}
\end{array}
\right\}
(-1)^{L-\frac{1}{2}-\sigma_{3}-\frac{1}{2}-\lambda_{2}-\frac{1}{2}-\lambda_{4}-1}.
\nonumber\\
\eeqa
\end{widetext}

Joining the resulting phase with the last three 3-J
coefficients of the equation Eq.~(\ref{V with six 3-jcoef}), the
following structure is obtained

\begin{widetext}
\beqa &\Rightarrow&\sum_{m_{3}\sigma_{3}m_{4}}
(-1)^{L-\frac{1}{2}-\sigma_{3}-\frac{1}{2}-\lambda_{2}-\frac{1}{2}-\lambda_{4}-1}
\left(\begin{array}{lll} l_{3} & \frac{1}{2} & j_{3}\\
m_{3} & \sigma_{3} & -\lambda_{3}\end{array}\right)
\left(\begin{array}{lll} l_{4} & \frac{1}{2} & j_{4}\\
m_{4} & \sigma_{3} & -\lambda_{4}\end{array}\right)
\left(\begin{array}{lll} l_{3} & l_{4} & L\\
-m_{3} & m_{4} & M\end{array}\right)\nonumber\\
&=&\sum_{m_{3}\sigma_{3}m_{4}}\left(\begin{array}{lll} j_{3} &l_{3}& \frac{1}{2}\\
-\lambda_{3} &m_{3}  & \sigma_{3}\end{array}\right)
\left(\begin{array}{lll} l_{4} & j_{4} &\frac{1}{2} \\
-m_{4} & \lambda_{4} & -\sigma_{3}\end{array}\right)
\left(\begin{array}{lll} l_{4} & l_{3} & L\\
m_{4} & -m_{3} & M\end{array}\right)
(-1)^{l_{3}+\frac{1}{2}+l_{4}+m_{4}+m_{3}-\sigma_{3}}\nonumber\\
&&(-1)^{L-\frac{1}{2}-\sigma_{3}-\frac{1}{2}-\lambda_{2}-\frac{1}{2}-\lambda_{4}-1
+l_{4}+l_{3}+L-l_{3}-\frac{1}{2}-l_{4}-m_{4}-m_{3}+\sigma_{3}}\nonumber\\
&=&\left(\begin{array}{lll} j_{3} & j_{4}& L \\
-\lambda_{3} & \lambda_{4} & M\end{array}\right)
\left\{\begin{array}{lll} j_{3} & j_{4}& L \\
l_{4} & l_{3} & \frac{1}{2}\end{array}\right\}
(-1)^{M}(-1)^{\frac{1}{2}+\lambda_{2}+\frac{1}{2}+\lambda_{4}},
\eeqa
where the 3-J coefficients can be written as

\beqa
\left(\begin{array}{lll} j_{1} & j_{2}& L \\
-\lambda_{1} & \lambda_{2} &
-M\end{array}\right)&=&\frac{(-1)^{j_{1}-j_{2}+M+j_{1}+j_{2}-L}}{\sqrt{2L+1}}
\langle j_{1}\lambda_{1},j_{2}-\lambda_{2}|L-M\rangle
\nonumber\\
\left(\begin{array}{lll} j_{3} & j_{4}& L \\
-\lambda_{3} & \lambda_{4} &
M\end{array}\right)&=&\frac{(-1)^{j_{3}-j_{4}-M+j_{3}+j_{4}-L}}{\sqrt{2L+1}}
\langle j_{3}\lambda_{3},j_{4}-\lambda_{4}|LM\rangle.
\eeqa
\end{widetext}

Another contribution to the general phase comes from lowering the
spin and color indices of the annihilation operators, which gives
the additional phases

\beqa
(-1)^{j_{2}-\lambda_{2}+\chi_2}
(-1)^{j_{4}-\lambda_{4}+\chi_4}.
\eeqa
With this the general phase is given by

\beqa
&(-1)^{M}&(-1)^{\frac{1}{2}+\lambda_{2}+\frac{1}{2}+\lambda_{4}}
(-1)^{j_{2}-\lambda_{2}+\chi_2}
(-1)^{j_{4}-\lambda_{4}+\chi_4}\nonumber\\
&\times&(-1)^{j_{1}-j_{2}+M+j_{1}+j_{2}-L}
(-1)^{j_{3}-j_{4}-M+j_{3}+j_{4}-L}\nonumber\\
&=&(-1)^{\frac{1}{2}+j_{2}+\frac{1}{2}+j_{4}}(-1)^{M}
(-1)^{\chi_2+\chi_4},
\eeqa
where $(-1)^{\chi_k}$ are the $SU(3)$-phases of color, as defined in \cite{jutta}.

Finally, the product of the SU(2) color generators can be written
as follows:

\begin{widetext}
\beqa
({\bd T}_a)^{c_1}_{c_2} ({\bd T}^a)^{c_3}_{c_4} & = &
\langle (1,0)c_2, (1,1)a \mid (1,0)c_1\rangle_1 (-1)^{\chi_a}
\langle (1,0)c_4, (1,1){\bar a} \mid (1,0)c_3\rangle_1
\langle (1,0)\mid\mid\mid {\bd T} \mid\mid\mid (1,0)\rangle^2.
\eeqa
\end{widetext}
The factor $\langle (1,0)\mid\mid\mid {\bd T} \mid\mid\mid
(1,0)\rangle$ is a triple reduced matrix element and its value is
given by twice the second order Casimir operator of $SU(3)$
\cite{jutta} which is equal to $\sqrt{8}$. The bar over an index
refers to the conjugate component. The index 1 at the end of an
$SU(3)$ Clebsch-Gordan coefficient is the multiplicity of the
coupling. Using the symmetry properties of the $SU(3)$
Clebsch-Gordan coefficients \cite{jutta} and changing the index
$a$ by $C$, we arrive at the expression

\beqa
& ({\bd T}_C)^{c_1}_{c_2} ({\bd T}^C)^{c_3}_{c_4}  = &
\nonumber \\
& 3(-1)^{\chi_{c_2} + \chi_{c_4} + \chi_C}
\langle (1,0)c_1, (0.1) {\bar c}_2 \mid (1,1) C\rangle_1 &
\nonumber \\
& \langle (1,0)c_3, (0.1) {\bar c}_4 \mid (1,1) {\bar C}\rangle_1 &.
\eeqa

Finally, the potential term is given by

\begin{widetext}
\beqa
\bd{V}&=&
\sum_{N_i,j_{i},\lambda_i,l_i,c_{i},\alpha,f,\alpha^{\prime},f^{\prime},L,M,C}
\left[b^{\dagger}_{\alpha_{1}(N_{1},l_{1},\frac{1}{2})j_{1}\lambda_{1}c_{1}f}
b^{\alpha_{2},f}_{(N_{2},l_{2},\frac{1}{2})j_{2}-\lambda_{2}-c_{2}}
\langle{j_{1}\lambda_{1}},j_{2}-\lambda_{2}|L-M\rangle
\langle(1,0)c_{1},(0,1){\bar c}_{2}|(1,1)C\rangle\right]
\nonumber\\
&&\left[b^{\dagger}_{\alpha_{3}(N_{3},l_{3},\frac{1}{2})j_{3}\lambda_{3}c_{3}f^{\prime}}
b^{\alpha_{4},f^{\prime}}_{(N_{4},l_{4},\frac{1}{2})j_{4}-\lambda_{4}-c_{4}}
\langle{j_{3}\lambda_{3}},j_{4}-\lambda_{4}|LM\rangle
\langle(1,0)c_{3},(0,1){\bar c}_{4}|(1,1){\bar C}\rangle\right]
(-1)^{M+\chi_C}
\nonumber\\
&&(-1)^{\frac{1}{2}+j_{2}+\frac{1}{2}+j_{4}}\frac{3}{2}\frac{1}{2L+1}
\int |\bd{x}|^{2}d|\bd{x}||\bd{y}|^{2}d|\bd{y}|
R_{N_{1}l_{1}}^{*}(|\bd{x}|) R_{N_{2}l_{2}}(|\bd{x}|)
R_{N_{3}l_{3}}^{*}(|\bd{y}|) R_{N_{4}l_{4}}(|\bd{y}|)
\nonumber\\
&&\int^{1}_{-1}d(cos\theta)P_{L}(cos\theta)V(|\bd{x}-\bd{y}|)
\nonumber\\
&& \prod_{i=1}^{4}\sqrt{(2l_{i}+1)(2j_{i}+1)}
\langle{l_{1}0},l_{2}0|L0\rangle \langle{l_{3}}0,l_{4}0|L0\rangle
\left\{\begin{array}{lll} j_{1} & l_{1}& \frac{1}{2} \\
l_{2} & j_{2} &L\end{array}\right\}
\left\{\begin{array}{lll} j_{3} & l_{3}& \frac{1}{2} \\
l_{4} & j_{4} &L\end{array}\right\}.
\eeqa
\end{widetext}

The Clebsch-Gordan and all recouping coefficients are calculated
numerically \cite{draayer1,draayer2}. In \cite{jutta} a useful
collection of formulas, comprising all symmetry relations known
for these coefficients, is given. In \cite{bahri1,bahri2} more
recent versions of the computer routines, for calculating the
$SU(3)$ Clebsch-Gordan and recoupling coefficients, are available.

As mentioned before, the potential term has the following form:

\begin{widetext}
\beqa {\bd V} & = &
\sum_{N_i,j_{i},\lambda_i,l_i,c_{i},\alpha,f,\alpha^{\prime},f^{\prime},L,M,C}
(-1)^{M}(-1)^{\chi_C}\nonumber\\
&&\left[\bd{b}^{\dagger}_{\alpha(N_1,l_{1},\frac{1}{2})j_{1}\lambda_{1},c_{1}f}
\bd{ b}^{\alpha,f}_{(N_2,l_{2},\frac{1}{2})j_{2}\lambda_{2},c_{2}}
\langle{j_{1}\lambda_{1},j_{2}\lambda_{2}|LM\rangle} \langle
(1,0)c_{1},(0,1)c_{2}|(1,1)C\rangle\right]
\nonumber \\
&&\left[\bd{
b}^{\dagger}_{\alpha^{\prime},(N_3,l_{3},\frac{1}{2})j_{3},\lambda_{3},c_{3}
f^{\prime}} \bd{
b}^{\alpha^{\prime},f^{\prime}}_{(N_4,l_{4},\frac{1}{2})j_{4}\lambda_{4},c_{4}}
\langle{j_{3}\lambda_{3},j_{4}\lambda_{4}|L-M\rangle} \langle
(1,0)c_{3},(0,1)c_{4}|(1,1){\bar C}\rangle\right]
\nonumber \\
&& \sqrt{(2j_1+1)(2j_3+1)}V(N_i,l_{i},j_{i},L)
\eeqa
\end{widetext}
The creation operator is in a (1,0) $SU(3)$-color  irrep and the
annihilation operator with a lower index is in a (0,1)
$SU(3)$-color irrep. A creation operator is coupled with the
annihilation operator to a (1,1) $SU(3)$-color irrep, denoting the
color generator.  The quantity $V(N_i,l_{i},j_{i},L)$ is the
intensity of each component of the interaction, and it reads \beqa
& & V(N_i,l_{i},j_{i},L) = \frac{3}{2}
 \frac{1}{(2L+1)}(-1)^{j_{2}+\frac{1}{2}+j_{4}+\frac{1}{2}}
 \nonumber\\
& & \times \int|\bd{x}|^{2}d|\bd{x}||\bd{y}|^{2}d|\bd{y}|
R^{*}_{N_1l_{1}}(|\bd{x}|) R_{N_2l_{2}}(|\bd{x}|)
\nonumber \\
&&R^{*}_{N_3l_{3}}(|\bd{y}|)R_{N_4l_{4}}(|\bd{y}|)
 \nonumber \\
& & \times  \int^{1}_{-1}d({\rm cos}\theta ) P_{L}({\rm
cos}\theta)V(|\bd{x}|,|\bd{y}|,{\rm cos}\theta)
\nonumber\\
& & \times
\frac{\prod^{4}_{i=1}\sqrt{(2l_{i}+1)(2j_{i}+1)}\langle{l_{1}0},l_{2}0|L0\rangle
\langle{l_{3}0},l_{4}0|L0\rangle}{\sqrt{(2j_1+1)(2j_3+1)}}
\nonumber \\
&&\left\{\begin{array}{lll} j_{1}&l_{1}&\frac{1}{2} \\
l_{2}&j_{2}&L\end{array}\right\}
\left\{\begin{array}{lll} j_{3}&l_{3}&\frac{1}{2} \\
l_{4}&j_{4}&L\end{array}\right\}. \nonumber \\
\label{v-2}
\eeqa

Taking into account that the only allowed value of L is L=0
and using a constant potential $V_{0}$ as residual
interaction, implies the following simplifications:

\beqa
& &\int^{1}_{-1}d({\rm cos}\theta ) P_{L}({\rm
cos}\theta)V(|\bd{x}|,|\bd{y}|,{\rm cos}\theta)=2V_0
\nonumber \\
& & \int|\bd{x}|^{2}d|\bd{x}| |\bd{y}|^{2}d|\bd{y}|
R^{*}_{N_1l_{1}}(|\bd{x}|)
R_{N_2l_{2}}(|\bd{x}|) \nonumber \\
&&
R^{*}_{N_3l_{3}}(|\bd{y}|)R_{N_4l_{4}}(|\bd{y}|)=\nonumber\\
& &=\int|\bd{x}|^{2}d|\bd{x}| R^{*}_{N_1l_{1}}(|\bd{x}|)
R_{N_2l_{2}}(|\bd{x}|)
\nonumber \\
&&\int|\bd{y}|^{2}d|\bd{y}|
R^{*}_{N_3l_{3}}(|\bd{y}|)R_{N_4l_{4}}(|\bd{y}|)=\nonumber\\
&&=\delta_{N_{1}N_{2}}\delta_{N_{3}N_{4}}\delta_{l_{1}l_{2}}\delta_{l_{3}l_{4}}\nonumber \\
&&
\langle{l_{1}0},l_{2}0|L0\rangle=\frac{(-1)^{l_{1}}}{\sqrt{2l_{1}+1}}\nonumber \\
&&\langle{l_{3}0},l_{4}0|L0\rangle=\frac{(-1)^{l_{3}}}{\sqrt{2l_{3}+1}}\nonumber \\
&&\left\{\begin{array}{lll} j_{1}&j_{2}&0 \nonumber \\
l_{1}&l_{2}&\frac{1}{2}\end{array}\right\}=\frac{(-1)^{j_{1}+l_{1}+\frac{1}{2}}}{\sqrt{(2j_{1}+1)(2l_{1}+1)}}\nonumber \\
&&\left\{\begin{array}{lll} j_{3}&j_{4}&0 \nonumber \\
l_{3}&l_{4}&\frac{1}{2}\end{array}\right\}=\frac{(-1)^{j_{3}+l_{3}+\frac{1}{2}}}{\sqrt{(2j_{3}+1)(2l_{3}+1)}}
\eeqa

therefore the intensities $V(N_i,l_{i},j_{i},L)$ are given by

\beqa
V(N_i,l_{i},j_{i},L)=\frac{V_{0}}{2}\delta_{L0}\delta_{l_{1}l_{2}}\delta_{l_{3}l_{4}}\delta_{j_{1}j_{2}}
\delta_{j_{3}j_{4}}\delta_{N_{1}N_{2}}\delta_{N_{3}N_{4}}. \eeqa

\section*{Appendix B: Constructing the kinetic energy operator:}

We start from the expression of the kinetic energy term obtained
in section II, without the mass term. The contribution of the mass term
will be considered Appendix C.

\begin{widetext}
\beqa \widetilde{\bd{K}}_j & = &
\sum_{N=j+\frac{1}{2}}^{\infty}
\sum_{N^\prime=N-1}^{N+1} \sum_{\lambda cf}
k_{jNN^{\prime}}\left[{\bd b}^{\dagger}_{\frac{1}{2}
(N,j+\frac{1}{2},\frac{1}{2})j\lambda cf} {\bd b}^{-\frac{1}{2}
(N^{\prime},j-\frac{1}{2},\frac{1}{2})j\lambda cf}+{\bd
b}^{\dagger}_{-\frac{1}{2} (N,j+\frac{1}{2},\frac{1}{2})j\lambda
cf} {\bd b}^{\frac{1}{2}
(N^{\prime},j-\frac{1}{2},\frac{1}{2})j\lambda cf}\right]
\nonumber \\
&& + \sum_{N=j+\frac{1}{2}}^{\infty}
\sum_{N^\prime=N-1}^{N+1} \sum_{\lambda cf}
k_{jNN^{\prime}}^{*}\left[{\bd b}^{\dagger}_{\frac{1}{2}
(N^{\prime},j-\frac{1}{2},\frac{1}{2})j\lambda cf} {\bd
b}^{-\frac{1}{2} (N,j+\frac{1}{2},\frac{1}{2})j\lambda cf}+{\bd
b}^{\dagger}_{-\frac{1}{2}
(N^{\prime},j-\frac{1}{2},\frac{1}{2})j\lambda cf} {\bd
b}^{\frac{1}{2} (N,j+\frac{1}{2},\frac{1}{2})j\lambda cf}\right]
\eeqa
\end{widetext}
where $N$ varies in steps of 2, and the $k_{jNN^{\prime}}$
constant is given by

\beqa
&k_{jNN^{\prime}}=&\nonumber\\
&i\int{r^{2}dr}\left[R^{*}_{N(j+\frac{1}{2})}(r)
\left(\frac{d}{dr}-\frac{j-\frac{1}{2}}{r}\right)R_{N^{\prime}(j-\frac{1}{2})}(r)\right]&. \eeqa

We use the 3-dimensional harmonic oscillator, i.e.,
($R_{n^{\prime}l}(r)=\textit{N}_{n^{\prime}l}\exp(-\frac{\gamma
r^{2}}{2})r^{l}L_{n^{\prime}}^{l+\frac{1}{2}}(\gamma r^{2})$) as
radial functions and the following convention:
$N^{\prime}=\frac{n^{\prime}-l}{2}$, which is the conventional
notation for the harmonic oscillator (at the end we will come back
to the $Nj$-notation, where $l=j-\frac{1}{2}$ and
$l^{\prime}=j+\frac{1}{2}$). The factors $k_{jNN^{\prime}}$ can be
written now as

\beqa K_{lnn^{\prime}}=i\int{r^{2}}dr
\left[R^{*}_{nl^{\prime}}(r)\left(
\frac{d}{dr}-\frac{l}{r}\right)R_{n^{\prime}l}(r)\right] \eeqa

The result of applying of the operator
$\left(\frac{d}{dr}-\frac{l}{r}\right)$ on the radial function is

\beqa
& \left(\frac{d}{dr}-\frac{l}{r}\right)R_{n^{\prime}l}(r) = &
\nonumber \\
& \textit{N}_{n^{\prime}l}
\left\{r^{l}\exp(-\frac{\gamma
r^{2}}{2})\frac{d}{dr}L_{n^{\prime}}^{l+\frac{1}{2}}(\gamma r^{2}) \right.&
\nonumber \\
& \left. -\gamma r^{l+1}\exp(-\frac{\gamma
r^{2}}{2})L_{n^{\prime}}^{l+\frac{1}{2}}(\gamma
r^{2})\right\}{\label{radial op1}} & .
\eeqa

From the recurrence relations of the Laguerre polynomials \beqa
L_{n}^{\alpha}(x)&=&L_{n}^{\alpha+1}(x)-L_{n-1}^{\alpha+1}(x)\nonumber\\
\frac{d}{dx}L_{n}^{\alpha}&=&-L_{n-1}^{\alpha+1}(x)
\eeqa

the first relation the equation Eq.~(\ref{radial op1}) can be
written as

\beqa
& \left(\frac{d}{dr}-\frac{l}{r}\right)R_{n^{\prime}l}(r)=&
\nonumber \\
& \textit{N}_{n^{\prime}l}\left\{r^{l}\exp(-\frac{\gamma
r^{2}}{2})\frac{d}{dr}L_{n^{\prime}}^{l+\frac{1}{2}}(\gamma r^{2})
\right. & \nonumber\\
&\left.-\gamma r^{l+1}\exp(-\frac{\gamma
r^{2}}{2})(L_{n^{\prime}}^{l+1+\frac{1}{2}}(\gamma
r^{2})-L_{n^{\prime}-1}^{l+1+\frac{1}{2}}(\gamma r^{2}))\right\} &
\nonumber \\
\eeqa and \beqa & r^{l}\exp(-\frac{\gamma
r^{2}}{2})\frac{d}{dr}L_{n^{\prime}}^{l+\frac{1}{2}}(\gamma
r^{2})= & \nonumber \\
& r^{l}\exp(-\frac{\gamma r^{2}}{2})(-2\gamma
r)L_{n^{\prime}-1}^{l+1+\frac{1}{2}}(x) &.
\nonumber \\
\eeqa
We obtain the final
expression

\beqa
& \left(\frac{d}{dr}-\frac{l}{r}\right) R_{n^{\prime}l}(r) = &
\nonumber \\
& \textit{N}_{n^{\prime}l}\left\{-2\gamma r^{l+1}\exp{(-\frac{\gamma
r^{2}}{2})}L_{n^{\prime}-1}^{l+1+\frac{1}{2}}(\gamma
r^{2})\right. & \nonumber\\
& \left.-\gamma r^{l+1}\exp{(-\frac{\gamma
r^{2}}{2})}L_{n^{\prime}}^{l+1+\frac{1}{2}}(\gamma
r^{2})\right. & \nonumber\\
&\left.+\gamma r^{l+1}\exp{(-\frac{\gamma
r^{2}}{2})}L_{n^{\prime}-1}^{l+1+\frac{1}{2}}(\gamma
r^{2})\right\} & \nonumber\\
&=\textit{N}_{n^{\prime}l}
\left\{-\gamma\frac{R_{n^{\prime}-1,l+1}(r)}{\textit{N}_{n^{\prime}-1,l+1}}
-\gamma\frac{R_{n^{\prime},l+1}(r)}{\textit{N}_{n^{\prime},l+1}}\right\}. &
\nonumber\\
\eeqa

Therefore, the connections for the harmonic oscillator are given
by \beqa k_{lnn^{\prime}}=
-i\gamma\frac{\textit{N}_{n^{\prime},l}}{\textit{N}_{n^{\prime}-1,l+1}}
\delta_{n,n^{\prime}-1}\delta_{l^{\prime},l+1}
-i\gamma\frac{\textit{N}_{n^{\prime},l}}{\textit{N}_{n^{\prime},l+1}}
\delta_{n,n^{\prime}}\delta_{l^{\prime},l+1}. \nonumber
\\\label{klnnp}
 \eeqa

In order to translate this expressions to the original notation we
write $n^{\prime}=\frac{N^{\prime}-l}{2}$ and
$n=\frac{N-l^{\prime}}{2}$, so for the first and second term of
 Eq.~\ref{klnnp} one has the following selection rules

\beqa
n&=&n^{\prime}-1\Rightarrow N^{\prime}=N+1 \nonumber\\
n&=&n^{\prime}\Rightarrow N^{\prime}=N-1.
\eeqa

The quotient to the normalization constants is given by

\beqa
& \frac{\textit{N}_{n^{\prime},l}}{\textit{N}_{n^{\prime}-1,l+1}}\delta_{n,n^{\prime}-1}\delta_{l^{\prime},l+1} = & \nonumber \\
& \left[\frac{2(n^{\prime})!}{\Gamma(n^{\prime}+l+\frac{3}{2})}\right]^{\frac{1}{2}}\gamma^{\frac{3}{4}+\frac{l}{2}}
\left[\frac{\Gamma(n^{\prime}-1+l+1+\frac{3}{2})}{2(n^{\prime}-1)!}\right]^{\frac{1}{2}} &
\frac{1}{\gamma^{\frac{3}{4}+\frac{l+1}{2}}} \nonumber\\
&=\sqrt{\frac{N-j+\frac{3}{2}}{2\gamma}} \delta_{N^\prime,N+1} &
\eeqa

and

\beqa
& \frac{\textit{N}_{n^{\prime},l}}{\textit{N}_{n^{\prime},l+1}}\delta_{n,n^{\prime}}\delta_{l^{\prime},l+1} = & \nonumber \\
& \left[\frac{2(n^{\prime})!}{\Gamma(n^{\prime}+l+\frac{3}{2})}\right]^{\frac{1}{2}}\gamma^{\frac{3}{4}+\frac{l}{2}}
\left[\frac{\Gamma(n^{\prime}+l+1+\frac{3}{2})}{2(n^{\prime})!}\right]^{\frac{1}{2}}
\frac{1}{\gamma^{\frac{3}{4}+\frac{l+1}{2}}} & \nonumber \\
&=\sqrt{\frac{N+j+\frac{3}{2}}{2\gamma}} \delta_{N^\prime,N-1} &.
 \eeqa 
 The kinetic energy term is written as ${\bd K}={\bd
K}^{(1+2)}+{\bd K}^{(3+4)}$ with
\begin{widetext}
\beqa {\bd K}^{(1+2)}&=&\sum_j \sum_{N=j+\frac{1}{2}}^{\infty}
\sum_{N^\prime=N-1}^{N+1}  \sum_{\lambda cf}
\left[-i\sqrt{\gamma}\sqrt{\frac{N-j+\frac{3}{2}}{2}}\delta_{N^{\prime},N+1}
-i\sqrt{\gamma}\sqrt{\frac{N+j+\frac{3}{2}}{2}}\delta_{N^{\prime},N-1}\right]\nonumber\\
&&\left({\bd
b}^{\dagger}_{\frac{1}{2}(N,j+\frac{1}{2},\frac{1}{2})j\lambda
cf}{\bd
b}^{-\frac{1}{2}(N^{\prime},j-\frac{1}{2},\frac{1}{2})j\lambda{cf}}+
{\bd
b}^{\dagger}_{-\frac{1}{2}(N,j+\frac{1}{2},\frac{1}{2})j\lambda
cf}{\bd
b}^{\frac{1}{2}(N^{\prime},j-\frac{1}{2},\frac{1}{2})j\lambda{cf}}\right)
\eeqa
\end{widetext}

\begin{widetext}
\beqa {\bd K}^{(3+4)}&=&\sum_j \sum_{N=j+\frac{1}{2}}^{\infty}
\sum_{N^\prime=N-1}^{N+1} \sum_{\lambda cf}
\left[i\sqrt{\gamma}\sqrt{\frac{N-j+\frac{3}{2}}{2}}\delta_{N^{\prime},N+1}
+i\sqrt{\gamma}\sqrt{\frac{N+j+\frac{3}{2}}{2}}\delta_{N^{\prime},N-1}\right]\nonumber\\
&&\left({\bd
b}^{\dagger}_{\frac{1}{2}(N^{\prime},j-\frac{1}{2},\frac{1}{2})j\lambda
cf}{\bd
b}^{-\frac{1}{2}(N,j+\frac{1}{2},\frac{1}{2})j\lambda{cf}}+ {\bd
b}^{\dagger}_{-\frac{1}{2}(N^{\prime},j-\frac{1}{2},\frac{1}{2})j\lambda
cf}{\bd
b}^{\frac{1}{2}(N,j+\frac{1}{2},\frac{1}{2})j\lambda{cf}}\right)
\eeqa
\end{widetext}

Finally, the operators ${\widetilde {\bd K}}_{+}=K^{(1)}+K^{(3)}$ and
${\widetilde {\bd K}}_{-}=K^{(2)}+K^{(4)}$ have the structure

\begin{widetext}
\beqa {\widetilde {\bd K}}_{+}&=&(-i\sqrt{\gamma})\sum_j
\sum_{N=j+\frac{1}{2}}^{\infty}
\sum_{N^\prime=N+1}^{N-1} \sum_{\lambda cf}
\left[\sqrt{\frac{N-j+\frac{3}{2}}{2}}\delta_{N^{\prime},N+1}
+\sqrt{\frac{N+j+\frac{3}{2}}{2}}\delta_{N^{\prime},N-1}\right]\nonumber\\
&&\left({\bd
b}^{\dagger}_{\frac{1}{2}(N,j+\frac{1}{2},\frac{1}{2})j\lambda
cf}{\bd
b}^{-\frac{1}{2}(N^{\prime},j-\frac{1}{2},\frac{1}{2})j\lambda{cf}}
-{\bd
b}^{\dagger}_{\frac{1}{2}(N^{\prime},j-\frac{1}{2},\frac{1}{2})j\lambda
cf}{\bd
b}^{-\frac{1}{2}(N,j+\frac{1}{2},\frac{1}{2})j\lambda{cf}}\right)
\nonumber\\
{\widetilde {\bd K}}_{-}&=&(-i\sqrt{\gamma})\sum_j
\sum_{N=j+\frac{1}{2}}^{\infty}
\sum_{N^\prime=N-1}^{N+1} \sum_{\lambda cf}
\left[\sqrt{\frac{N-j+\frac{3}{2}}{2}}\delta_{N^{\prime},N+1}
+\sqrt{\frac{N+j+\frac{3}{2}}{2}}\delta_{N^{\prime},N-1}\right]\nonumber\\
&&\left({\bd
b}^{\dagger}_{-\frac{1}{2}(N,j+\frac{1}{2},\frac{1}{2})j\lambda
cf}{\bd
b}^{\frac{1}{2}(N^{\prime},j-\frac{1}{2},\frac{1}{2})j\lambda{cf}}
-{\bd
b}^{\dagger}_{-\frac{1}{2}(N^{\prime},j-\frac{1}{2},\frac{1}{2})j\lambda
cf}{\bd
b}^{\frac{1}{2}(N,j+\frac{1}{2},\frac{1}{2})j\lambda{cf}}\right)
\eeqa
\end{widetext}

Now we redefine, for each set $N,j$ with $l=j+\frac{1}{2}$, the
fermion creation operators multiplying them by $(-i)$ and the
corresponding annihilation operator by $(+i)$. In this way, the
anti-commutation relations are preserved. Explicitly, the mapping
is given by

\beqa (-i){\bd b}^{\dagger}_{\pm\frac{1}{2}
(N,j+\frac{1}{2},\frac{1}{2})j\lambda cf} & \rightarrow &
{\bd b}^{\dagger}_{\pm\frac{1}{2}
(N,j+\frac{1}{2},\frac{1}{2})j\lambda cf}
\nonumber \\
(i){\bd b}^{\pm\frac{1}{2} (N,j+\frac{1}{2},\frac{1}{2})j\lambda cf}
& \rightarrow & {\bd b}^{\pm\frac{1}{2}
(N,j+\frac{1}{2},\frac{1}{2})j\lambda cf}.  
\nonumber \\
\label{renaming}
\eeqa
With this the
kinetic energy parts acquire the form

\begin{widetext}
\beqa {\widetilde {\bd K}}_{+}&=&(\sqrt{\gamma})\sum_j
\sum_{N=j+\frac{1}{2}}^{\infty}
\sum_{N^\prime=N-1}^{N+1} \sum_{\lambda cf}
\left[\sqrt{\frac{N-j+\frac{3}{2}}{2}}\delta_{N^{\prime},N+1}
+\sqrt{\frac{N+j+\frac{3}{2}}{2}}\delta_{N^{\prime},N-1}\right]
\nonumber\\
&&\left({\bd
b}^{\dagger}_{\frac{1}{2}(N,j+\frac{1}{2},\frac{1}{2})j\lambda
cf}{\bd
b}^{-\frac{1}{2}(N^{\prime},j-\frac{1}{2},\frac{1}{2})j\lambda{cf}}
+{\bd
b}^{\dagger}_{\frac{1}{2}(N^{\prime},j-\frac{1}{2},\frac{1}{2})j\lambda
cf}{\bd
b}^{-\frac{1}{2}(N,j+\frac{1}{2},\frac{1}{2})j\lambda{cf}}\right)
\nonumber\\
{\widetilde {\bd K}}_{-}&=&(\sqrt{\gamma})\sum_j
\sum_{N=j+\frac{1}{2}}^{\infty}
\sum_{N^\prime=N-1}^{N+1} \sum_{\lambda cf}
\left[\sqrt{\frac{N-j+\frac{3}{2}}{2}}\delta_{N^{\prime},N+1}
+\sqrt{\frac{N+j+\frac{3}{2}}{2}}\delta_{N^{\prime},N-1}\right]\nonumber\\
&&\left({\bd
b}^{\dagger}_{-\frac{1}{2}(N,j+\frac{1}{2},\frac{1}{2})j\lambda
cf}{\bd
b}^{\frac{1}{2}(N^{\prime},j-\frac{1}{2},\frac{1}{2})j\lambda{cf}}
+{\bd
b}^{\dagger}_{-\frac{1}{2}(N^{\prime},j-\frac{1}{2},\frac{1}{2})j\lambda
cf}{\bd
b}^{\frac{1}{2}(N,j+\frac{1}{2},\frac{1}{2})j\lambda{cf}}\right)
\eeqa
\end{widetext}

\section*{Appendix C: Contribution of the mass term}

Substituting the fermion fields into the expression of the mass
term, we obtain

\begin{widetext}
\beqa &&\int{d{\bd x}}{\bd \psi}^{\dagger}(\bd{x})\beta{m_{0}}{\bd
\psi}(\bd{x})=
\int{d{\bd x}}\left({\bd
\psi}^{\dagger}_{1}(\bd{x}),{\bd
\psi}^{\dagger}_{2}(\bd{x})\right)
\left(\begin{array}{ll} m_{0}\bd{1} & 0 \\
0 &  -m_{0}\bd{1} \end{array}\right)
\left(\begin{array}{l}\bd{\psi}_{1}(\bd{x})\\
\bd{\psi}_{2}(\bd{x})\end{array}\right)\nonumber\\
&&=m_{0}\int{d{\bd x}}\left({\bd \psi}^{\dagger}_{1}(\bd{x}){\bd
\psi}_{1}(\bd{x})
-{\bd \psi}^{\dagger}_{2}(\bd{x}){\bd \psi}_{2}(\bd{x})\right)\nonumber\\
&&=m_{0}\int{d{\bd
x}}\sum_{N_{1}N_{3}l_{1}l_{3}m_{1}m_{3}j_{1}j_{3}\lambda_{1}\lambda_{3}\sigma_{1}\sigma_{3}cf}
\langle{l_{1}}m_{1},\frac{1}{2}\sigma_{1}|j_{1}\lambda_{1}\rangle
\langle{l_{3}}m_{3},\frac{1}{2}\sigma_{3}|j_{3}\lambda_{3}\rangle\nonumber\\
&&R_{N_{1}l_{1}}^{*}(|\bd{x}|)R_{N_{3}l_{3}}(|\bd{x}|)
Y_{l_{1}m_{1}}^{*}(\Omega_{x})Y_{l_{3}m_{3}}(\Omega_{x})
\chi^{*}_{\sigma_{1}}\chi_{\sigma_{3}} {\bd
b}^{\dagger}_{\frac{1}{2}(N_{1}l_{1}\frac{1}{2})j_{1}\lambda_{1}cf}
{\bd b}^{\frac{1}{2}(N_{3}l_{3}\frac{1}{2})j_{3}\lambda_{3}cf}\nonumber\\
&& -\int{d{\bd
x}}\sum_{N_{2}N_{4}l_{2}l_{4}m_{2}m_{4}j_{2}j_{4}\lambda_{2}\lambda_{4}\sigma_{2}\sigma_{4}cf}
\langle{l_{2}}m_{2},\frac{1}{2}\sigma_{2}|j_{2}\lambda_{2}\rangle
\langle{l_{4}}m_{4},\frac{1}{2}\sigma_{4}|j_{4}\lambda_{4}\rangle\nonumber\\
&&R_{N_{2}l_{2}}^{*}(|\bd{x}|)R_{N_{4}l_{4}}(|\bd{x}|)
Y_{l_{2}m_{2}}^{*}(\Omega_{x})Y_{l_{4}m_{4}}(\Omega_{x})
\chi^{*}_{\sigma_{2}}\chi_{\sigma_{4}} {\bd
b}^{\dagger}_{-\frac{1}{2}(N_{2}l_{2}\frac{1}{2})j_{2}\lambda_{2}cf}
{\bd b}^{-\frac{1}{2}(N_{4}l_{4}\frac{1}{2})j_{4}\lambda_{4}cf}.
 \eeqa
\end{widetext}
The radial and the angular integral leads to the restriction
$\delta_{N_{1}N_{3}}\delta_{l_{1}l_{3}}\delta_{m_{1}m_{3}}\delta_{\sigma_{1}\sigma_{3}}$
and
$\delta_{N_{2}N_{4}}\delta_{l_{2}l_{4}}\delta_{m_{2}m_{4}}\delta_{\sigma_{2}\sigma_{4}}$,
respectively. Thus, the mass term can be written as (applying also the
mapping of the fermion creation and annihilation operators, as given
in Eq.~(\ref{renaming})):

\begin{widetext}
\beqa \int{d{\bd x}}{\bd \psi}^{\dagger}(\bd{x})\beta{m_{0}}{\bd
\psi}(\bd{x}) & = & m_{0}\sum_{Nlj\lambda{cf}} \left\{{\bd
b}^{\dagger}_{\frac{1}{2}(Nl\frac{1}{2})j\lambda{cf}} {\bd
b}^{\frac{1}{2}(Nl\frac{1}{2})j\lambda{cf}} -{\bd
b}^{\dagger}_{-\frac{1}{2}N(l\frac{1}{2})j\lambda{cf}} {\bd
b}^{-\frac{1}{2}N(l\frac{1}{2})j\lambda{cf}}\right\}
\nonumber \\
& = &
m_{0}\sum_{j}\sum_{\lambda{cf}}
\sum_{N^{\prime}=j-\frac{1}{2}}^{\infty} \left({\bd
b}^{\dagger}_{\frac{1}{2}(N^{\prime},j-\frac{1}{2},\frac{1}{2})j\lambda{cf}}
{\bd
b}^{\frac{1}{2}(N^{\prime},j-\frac{1}{2},\frac{1}{2})j\lambda{cf}}
-{\bd
b}^{\dagger}_{-\frac{1}{2}(N^{\prime},j-\frac{1}{2},\frac{1}{2})j\lambda{cf}}
{\bd
b}^{-\frac{1}{2}(N^{\prime},j-\frac{1}{2},\frac{1}{2})j\lambda{cf}}\right)
\nonumber\\
&&+m_{0}\sum_{j}\sum_{\lambda{cf}}\sum_{N=j+\frac{1}{2}}^{\infty}
\left({\bd
b}^{\dagger}_{\frac{1}{2}(N,j+\frac{1}{2},\frac{1}{2})j\lambda{cf}}
{\bd
b}^{\frac{1}{2}(N,j+\frac{1}{2},\frac{1}{2})j\lambda{cf}}
-{\bd
b}^{\dagger}_{-\frac{1}{2}(N,j+\frac{1}{2},\frac{1}{2})j\lambda{cf}}
{\bd
b}^{-\frac{1}{2}(N,j+\frac{1}{2},\frac{1}{2})j\lambda{cf}}\right),
\eeqa
\end{widetext}
where $N$ and $N^\prime$ vary in steps of 2. Now, we apply the
same unitary transformation as described in the text, i.e.,

\beqa
{\bd b}^{\dagger}_{\pm\frac{1}{2}(N,j+\frac{1}{2},\frac{1}{2})j\lambda{cf}}&=&
\sum_{k}\alpha^{*}_{Nk}
\widehat{{\bd b}}^{\dagger}_{\pm\frac{1}{2}(k,j+\frac{1}{2},\frac{1}{2})j\lambda cf}
\nonumber\\
{\bd
b}^{\dagger}_{\pm\frac{1}{2}(N^{\prime},j-\frac{1}{2},\frac{1}{2})\frac{1}{2}\lambda{cf}}&=&
\sum_{q}\beta^{*}_{N^{\prime}q} \widehat{{\bd
b}}^{\dagger}_{\pm\frac{1}{2}(q,j-\frac{1}{2},\frac{1}{2})j\lambda
cf} .
 \eeqa 
 With this, the mass term is written

\begin{widetext}
\beqa
\int{d{\bd x}}{\bd \psi}^{\dagger}(\bd{x})\beta{m_{0}}{\bd \psi}(\bd{x})&=&
\nonumber\\
&=&\sum_{j}\sum_{\lambda{cf}}\sum_{q}\left\{m_{0,q,j-\frac{1}{2}}
\left(\widehat{{\bd
b}}^{\dagger}_{\frac{1}{2}(q,j-\frac{1}{2},\frac{1}{2})j\lambda
cf} \widehat{{\bd
b}}^{\frac{1}{2}(q,j-\frac{1}{2},\frac{1}{2})j\lambda cf}
-\widehat{{\bd
b}}^{\dagger}_{-\frac{1}{2}(q,j-\frac{1}{2},\frac{1}{2})j\lambda
cf} \widehat{{\bd
b}}^{-\frac{1}{2}(q,j-\frac{1}{2},\frac{1}{2})j\lambda
cf}\right)\right\}
\nonumber\\
&+&\sum_{j}\sum_{\lambda{cf}}\sum_{k} \left\{
m_{0,k,j+\frac{1}{2}}\left(\widehat{{\bd
b}}^{\dagger}_{\frac{1}{2}(k,j+\frac{1}{2},\frac{1}{2})j\lambda
cf} \widehat{{\bd
b}}^{\frac{1}{2}(k,j+\frac{1}{2},\frac{1}{2})j\lambda cf}
-\widehat{{\bd
b}}^{\dagger}_{-\frac{1}{2}(k,j+\frac{1}{2},\frac{1}{2})j\lambda
cf} \widehat{{\bd
b}}^{-\frac{1}{2}(k,j+\frac{1}{2},\frac{1}{2})j\lambda
cf}\right)\right\}.
\nonumber \\
\eeqa
\end{widetext}
Due to the method described in section III the only remaining
terms correspond to k=q. With this the masses become

\beqa
m_{0,k,j-\frac{1}{2}}&=&
\sum_{N^{\prime}=j-\frac{1}{2}}^{n-1}m_{0}|\beta_{jN^{\prime}k}|^{2}\nonumber\\
m_{0,k,j+\frac{1}{2}}&=&
\sum_{N=j+\frac{1}{2}}^{n}m_{0}|\alpha_{jNk}|^{2}.{\label{m_{0,s,p}-2}}
\eeqa Finally, the mass contribution to the BCS kinetic energy is
given by

\begin{widetext}
\beqa
&{\bd K}&^{BCS}_{masa}=\nonumber\\
&&=\sum_{\lambda{cf}}\sum_{kj}
\left\{\left[m_{0,k,j-\frac{1}{2}}c_{j+\frac{1}{2},k}^{2}-m_{0,k,j+\frac{1}{2}}s_{j+\frac{1}{2},k}^{2}\right]
b^{\dagger}_{j-\frac{1}{2}(kj)\lambda{cf}}b_{j-\frac{1}{2}}^{(kj)\lambda{cf}}\right.
\nonumber\\
&&+\left.\left[m_{0,k,j-\frac{1}{2}}c_{j-\frac{1}{2},k}^{2}-m_{0,k,j+\frac{1}{2}}s_{j-\frac{1}{2},k}^{2}\right]
d^{\dagger(kj)\lambda{cf}}_{j-\frac{1}{2}}d_{j-\frac{1}{2}(kj)\lambda{cf}}\right.
\nonumber\\
&&+\left.\left[m_{0,k,j+\frac{1}{2}}c_{j-\frac{1}{2},k}^{2}-m_{0,k,j-\frac{1}{2}}s_{j-\frac{1}{2},k}^{2}\right]
b^{\dagger}_{j+\frac{1}{2}(kj)\lambda{cf}}b_{j+\frac{1}{2}}^{(kj)\lambda{cf}}\right.
\nonumber\\
&&+\left.\left[m_{0,k,j+\frac{1}{2}}c_{j+\frac{1}{2},k}^{2}-m_{0,k,j-\frac{1}{2}}s_{j+\frac{1}{2},k}^{2}\right]
d^{\dagger(kj)\lambda{cf}}_{j+\frac{1}{2}}d_{j+\frac{1}{2}(kj)\lambda{cf}}\right.
\nonumber\\
&&+\left. 12 \left[
m_{0,k,j-\frac{1}{2}}(s_{j+\frac{1}{2},k}^{2}-c_{j-\frac{1}{2},k}^{2})
+m_{0,k,j+\frac{1}{2}}(s_{j-\frac{1}{2},k}^{2}-c_{j+\frac{1}{2},k}^{2})\right]\right.
\nonumber\\
&&-\left.\left[
(m_{0,k,j-\frac{1}{2}}+m_{0,k,j+\frac{1}{2}})s_{j+\frac{1}{2},k}c_{j+\frac{1}{2},k}\right]
\left[b^{\dagger}_{j-\frac{1}{2}(kj)\lambda{cf}}d^{\dagger(kj)\lambda{cf}}_{j+\frac{1}{2}}
+d_{j+\frac{1}{2}(kj)\lambda{cf}}b^{(kj)\lambda{cf}}_{j-\frac{1}{2}}\right]\right.
\nonumber\\
&&-\left.\left[
(m_{0,k,j-\frac{1}{2}}+m_{0,k,j+\frac{1}{2}})s_{j-\frac{1}{2},k}c_{j-\frac{1}{2},k}\right]
\left[b^{\dagger}_{j+\frac{1}{2}(kj)\lambda{cf}}d^{\dagger(kj)\lambda{cf}}_{j-\frac{1}{2}}
+d_{j-\frac{1}{2}(kj)\lambda{cf}}b^{(kj)\lambda{cf}}_{j+\frac{1}{2}}\right]\right\}\nonumber.
\eeqa
\end{widetext}

\section*{APPENDIX D: The kinetic energy term for three levels and
a fixed $N$ and $j$}

The Dirac picture of the three-energy levels implies six states,
three with positive energy and three with negative energy. Fixing
j and N, we are working in a determined column with total spin j
and the possible connections
$(N-1,j-\frac{1}{2})\longleftrightarrow(N,j+\frac{1}{2})$ and
$(N,j+\frac{1}{2})\longleftrightarrow(N+1,j-\frac{1}{2})$. The
operators ${\bd K}_{+}$ and ${\bd K}_{-}$  of this case have the
following structure, respectively,

\beqa & {\widetilde {\bd K}}_{+}=\sqrt{\gamma}&
\nonumber \\
& \left[\left(\frac{N-j+\frac{3}{2}}{2}\right)^{\frac{1}{2}}
b^{\dagger}_{\frac{1}{2}(N,j+\frac{1}{2},\frac{1}{2})j\lambda{cf}}
b^{-\frac{1}{2}(N+1,j-\frac{1}{2},\frac{1}{2})j\lambda{cf}}\right.&
\nonumber\\
&\left.+\left(\frac{N+j+\frac{3}{2}}{2}\right)^{\frac{1}{2}}
b^{\dagger}_{\frac{1}{2}(N,j+\frac{1}{2},\frac{1}{2})j\lambda{cf}}
b^{-\frac{1}{2}(N-1,j-\frac{1}{2},\frac{1}{2})j\lambda{cf}}\right.&
\nonumber\\
&\left.+\left(\frac{N-j+\frac{3}{2}}{2}\right)^{\frac{1}{2}}
b^{\dagger}_{\frac{1}{2}(N+1,j-\frac{1}{2},\frac{1}{2})j\lambda{cf}}
b^{-\frac{1}{2}(N,j+\frac{1}{2},\frac{1}{2})j\lambda{cf}}\right.&
\nonumber\\
&\left.+\left(\frac{N+j+\frac{3}{2}}{2}\right)^{\frac{1}{2}}
b^{\dagger}_{\frac{1}{2}(N-1,j-\frac{1}{2},\frac{1}{2})j\lambda{cf}}
b^{-\frac{1}{2}(N,j+\frac{1}{2},\frac{1}{2})j\lambda{cf}}\right]{\label{K_{+}^{N,j;fixed}}}
&
\nonumber \\
\eeqa

\beqa & {\widetilde {\bd K}}_{-}=\sqrt{\gamma} &
\nonumber \\
& \left[\left(\frac{N-j+\frac{3}{2}}{2}\right)^{\frac{1}{2}}
b^{\dagger}_{-\frac{1}{2}(N,j+\frac{1}{2},\frac{1}{2})j\lambda{cf}}
b^{\frac{1}{2}(N+1,j-\frac{1}{2},\frac{1}{2})j\lambda{cf}}\right.&
\nonumber\\
&\left.+\left(\frac{N+j+\frac{3}{2}}{2}\right)^{\frac{1}{2}}
b^{\dagger}_{-\frac{1}{2}(N,j+\frac{1}{2},\frac{1}{2})j\lambda{cf}}
b^{\frac{1}{2}(N-1,j-\frac{1}{2},\frac{1}{2})j\lambda{cf}}\right.&
\nonumber\\
&\left.+\left(\frac{N-j+\frac{3}{2}}{2}\right)^{\frac{1}{2}}
b^{\dagger}_{-\frac{1}{2}(N+1,j-\frac{1}{2},\frac{1}{2})j\lambda{cf}}
b^{\frac{1}{2}(N,j+\frac{1}{2},\frac{1}{2})j\lambda{cf}}\right.&
\nonumber\\
&\left.+\left(\frac{N+j+\frac{3}{2}}{2}\right)^{\frac{1}{2}}
b^{\dagger}_{-\frac{1}{2}(N-1,j-\frac{1}{2},\frac{1}{2})j\lambda{cf}}
b^{\frac{1}{2}(N,j+\frac{1}{2},\frac{1}{2})j\lambda{cf}}\right].{\label{K_{-}^{N,j;fixed}}}&
\nonumber \\
\eeqa

The commutation relation of these operators is given by (omitting
the sub-indices $j \lambda c f$ for simplicity)

\begin{widetext}

\beqa [\widetilde{\bd{K}}_{+},\widetilde{\bd{K}}_{-}]&=&
\gamma\left\{\left(\frac{N-j+\frac{3}{2}}{2}\right)
\left[\bd{N}_{\frac{1}{2}(N,j+\frac{1}{2},\frac{1}{2})}-\bd{N}_{-\frac{1}{2}(N+1,j-\frac{1}{2},\frac{1}{2})}\right]\right.\nonumber\\
&&\left.-\left(\frac{N-j+\frac{3}{2}}{2}\right)^{\frac{1}{2}}\left(\frac{N+j+\frac{3}{2}}{2}\right)^{\frac{1}{2}}
\bd{b}^{\dagger}_{-\frac{1}{2}(N-1,j-\frac{1}{2},\frac{1}{2})}
\bd{b}^{-\frac{1}{2}(N+1,j-\frac{1}{2},\frac{1}{2})}\right.\nonumber\\
&&\left.-\left(\frac{N-j+\frac{3}{2}}{2}\right)^{\frac{1}{2}}\left(\frac{N+j+\frac{3}{2}}{2}\right)^{\frac{1}{2}}
\bd{b}^{\dagger}_{-\frac{1}{2}(N+1,j-\frac{1}{2},\frac{1}{2})}
\bd{b}^{-\frac{1}{2}(N-1,j-\frac{1}{2},\frac{1}{2})}\right.\nonumber\\
&&\left.+\left(\frac{N+j+\frac{3}{2}}{2}\right)
\left[\bd{N}_{\frac{1}{2}(N,j+\frac{1}{2},\frac{1}{2})}-\bd{N}_{-\frac{1}{2}(N-1,j-\frac{1}{2},\frac{1}{2})}\right]\right.\nonumber\\
&&\left.+\left(\frac{N-j+\frac{3}{2}}{2}\right)
\left[\bd{N}_{\frac{1}{2}(N+1,j-\frac{1}{2},\frac{1}{2})}-\bd{N}_{-\frac{1}{2}(N,j+\frac{1}{2},\frac{1}{2})}\right]\right.\nonumber\\
&&\left.+\left(\frac{N-j+\frac{3}{2}}{2}\right)^{\frac{1}{2}}\left(\frac{N+j+\frac{3}{2}}{2}\right)^{\frac{1}{2}}
\bd{b}^{\dagger}_{\frac{1}{2}(N+1,j-\frac{1}{2},\frac{1}{2})}
\bd{b}^{\frac{1}{2}(N-1,j-\frac{1}{2},\frac{1}{2})}\right.\nonumber\\
&&\left.+\left(\frac{N-j+\frac{3}{2}}{2}\right)^{\frac{1}{2}}\left(\frac{N+j+\frac{3}{2}}{2}\right)^{\frac{1}{2}}
\bd{b}^{\dagger}_{\frac{1}{2}(N-1,j-\frac{1}{2},\frac{1}{2})}
\bd{b}^{\frac{1}{2}(N+1,j-\frac{1}{2},\frac{1}{2})}\right.\nonumber\\
&&\left.+\left(\frac{N+j+\frac{3}{2}}{2}\right)
\left[\bd{N}_{\frac{1}{2}(N-1,j-\frac{1}{2},\frac{1}{2})}-\bd{N}_{-\frac{1}{2}(N,j+\frac{1}{2},\frac{1}{2})}\right]
\right\}\nonumber\\
&=&2\widetilde{\bd{K}}_{0}.{\label{[K+,K-]=2K_{0}}} \eeqa

\beqa
[&\widetilde{\bd{K}}_{0}&,\widetilde{\bd{K}}_{+}]=\nonumber\\
&=&\left(\frac{\gamma^{\frac{3}{2}}}{2}\right)
\left\{\left(\frac{N-j+\frac{3}{2}}{2}\right)^{\frac{3}{2}}
\bd{b}^{\dagger}_{\frac{1}{2}(N,j+\frac{1}{2},\frac{1}{2})}
\bd{b}^{-\frac{1}{2}(N+1,j-\frac{1}{2},\frac{1}{2})}\right.\nonumber\\
&&\left.+\left(\frac{N-j+\frac{3}{2}}{2}\right)\left(\frac{N+j+\frac{3}{2}}{2}\right)^{\frac{1}{2}}
\bd{b}^{\dagger}_{\frac{1}{2}(N,j+\frac{1}{2},\frac{1}{2})}
\bd{b}^{-\frac{1}{2}(N-1,j-\frac{1}{2},\frac{1}{2})}\right.\nonumber\\
&&\left.+\left(\frac{N-j+\frac{3}{2}}{2}\right)^{\frac{3}{2}}
\bd{b}^{\dagger}_{\frac{1}{2}(N,j+\frac{1}{2},\frac{1}{2})}
\bd{b}^{-\frac{1}{2}(N+1,j-\frac{1}{2},\frac{1}{2})}\right.\nonumber\\
&&\left.+\left(\frac{N+j+\frac{3}{2}}{2}\right)\left(\frac{N-j+\frac{3}{2}}{2}\right)^{\frac{1}{2}}
\bd{b}^{\dagger}_{\frac{1}{2}(N,j+\frac{1}{2},\frac{1}{2})}
\bd{b}^{-\frac{1}{2}(N+1,j-\frac{1}{2},\frac{1}{2})}\right.\nonumber\\
&&\left.+\left(\frac{N-j+\frac{3}{2}}{2}\right)\left(\frac{N+j+\frac{3}{2}}{2}\right)^{\frac{1}{2}}
\bd{b}^{\dagger}_{\frac{1}{2}(N,j+\frac{1}{2},\frac{1}{2})}
\bd{b}^{-\frac{1}{2}(N-1,j-\frac{1}{2},\frac{1}{2})}\right.\nonumber\\
&&\left.+\left(\frac{N+j+\frac{3}{2}}{2}\right)\left(\frac{N-j+\frac{3}{2}}{2}\right)^{\frac{1}{2}}
\bd{b}^{\dagger}_{\frac{1}{2}(N,j+\frac{1}{2},\frac{1}{2})}
\bd{b}^{-\frac{1}{2}(N+1,j-\frac{1}{2},\frac{1}{2})}\right.\nonumber\\
&&\left.+\left(\frac{N+j+\frac{3}{2}}{2}\right)^{\frac{3}{2}}
\bd{b}^{\dagger}_{\frac{1}{2}(N,j+\frac{1}{2},\frac{1}{2})}
\bd{b}^{-\frac{1}{2}(N-1,j-\frac{1}{2},\frac{1}{2})}\right.\nonumber\\
&&\left.+\left(\frac{N+j+\frac{3}{2}}{2}\right)^{\frac{3}{2}}
\bd{b}^{\dagger}_{\frac{1}{2}(N,j+\frac{1}{2},\frac{1}{2})}
\bd{b}^{-\frac{1}{2}(N-1,j-\frac{1}{2},\frac{1}{2})}\right.\nonumber\\
&&\left.+\left(\frac{N-j+\frac{3}{2}}{2}\right)^{\frac{3}{2}}
\bd{b}^{\dagger}_{\frac{1}{2}(N+1,j-\frac{1}{2},\frac{1}{2})}
\bd{b}^{-\frac{1}{2}(N,j+\frac{1}{2},\frac{1}{2})}\right.\nonumber\\
&&\left.+\left(\frac{N-j+\frac{3}{2}}{2}\right)^{\frac{3}{2}}
\bd{b}^{\dagger}_{\frac{1}{2}(N+1,j-\frac{1}{2},\frac{1}{2})}
\bd{b}^{-\frac{1}{2}(N,j+\frac{1}{2},\frac{1}{2})}\right.\nonumber\\
&&\left.+\left(\frac{N-j+\frac{3}{2}}{2}\right)\left(\frac{N+j+\frac{3}{2}}{2}\right)^{\frac{1}{2}}
\bd{b}^{\dagger}_{\frac{1}{2}(N-1,j-\frac{1}{2},\frac{1}{2})}
\bd{b}^{-\frac{1}{2}(N,j+\frac{1}{2},\frac{1}{2})}\right.\nonumber\\
&&\left.+\left(\frac{N+j+\frac{3}{2}}{2}\right)\left(\frac{N-j+\frac{3}{2}}{2}\right)^{\frac{1}{2}}
\bd{b}^{\dagger}_{\frac{1}{2}(N+1,j-\frac{1}{2},\frac{1}{2})}
\bd{b}^{-\frac{1}{2}(N,j+\frac{1}{2},\frac{1}{2})}\right.\nonumber\\
&&\left.+\left(\frac{N-j+\frac{3}{2}}{2}\right)\left(\frac{N+j+\frac{3}{2}}{2}\right)^{\frac{1}{2}}
\bd{b}^{\dagger}_{\frac{1}{2}(N-1,j-\frac{1}{2},\frac{1}{2})}
\bd{b}^{-\frac{1}{2}(N,j+\frac{1}{2},\frac{1}{2})}\right.\nonumber\\
&&\left.+\left(\frac{N+j+\frac{3}{2}}{2}\right)^{\frac{3}{2}}
\bd{b}^{\dagger}_{\frac{1}{2}(N-1,j-\frac{1}{2},\frac{1}{2})}
\bd{b}^{-\frac{1}{2}(N,j+\frac{1}{2},\frac{1}{2})}\right.\nonumber\\
&&\left.+\left(\frac{N+j+\frac{3}{2}}{2}\right)\left(\frac{N-j+\frac{3}{2}}{2}\right)^{\frac{1}{2}}
\bd{b}^{\dagger}_{\frac{1}{2}(N+1,j-\frac{1}{2},\frac{1}{2})}
\bd{b}^{-\frac{1}{2}(N,j+\frac{1}{2},\frac{1}{2})}\right.\nonumber\\
&&\left.+\left(\frac{N+j+\frac{3}{2}}{2}\right)^{\frac{3}{2}}
\bd{b}^{\dagger}_{\frac{1}{2}(N-1,j-\frac{1}{2},\frac{1}{2})}
\bd{b}^{-\frac{1}{2}(N,j+\frac{1}{2},\frac{1}{2})}
\right\}\nonumber\\
&=&\frac{\gamma}{2}\sqrt{\gamma} \left\{
\left[2\left(\frac{N-j+\frac{3}{2}}{2}\right)+2\left(\frac{N+j+\frac{3}{2}}{2}\right)\right]
\left(\frac{N-j+\frac{3}{2}}{2}\right)^{\frac{1}{2}}
\bd{b}^{\dagger}_{\frac{1}{2}(N,j+\frac{1}{2},\frac{1}{2})}
\bd{b}^{-\frac{1}{2}(N+1,j-\frac{1}{2},\frac{1}{2})}\right.\nonumber\\
&&\left.+\left[2\left(\frac{N-j+\frac{3}{2}}{2}\right)+2\left(\frac{N+j+\frac{3}{2}}{2}\right)\right]
\left(\frac{N+j+\frac{3}{2}}{2}\right)^{\frac{1}{2}}
\bd{b}^{\dagger}_{\frac{1}{2}(N,j+\frac{1}{2},\frac{1}{2})}
\bd{b}^{-\frac{1}{2}(N-1,j-\frac{1}{2},\frac{1}{2})}\right.\nonumber\\
&&\left.+\left[2\left(\frac{N-j+\frac{3}{2}}{2}\right)+2\left(\frac{N+j+\frac{3}{2}}{2}\right)\right]
\left(\frac{N-j+\frac{3}{2}}{2}\right)^{\frac{1}{2}}
\bd{b}^{\dagger}_{\frac{1}{2}(N+1,j-\frac{1}{2},\frac{1}{2})}
\bd{b}^{-\frac{1}{2}(N,j+\frac{1}{2},\frac{1}{2})}\right.\nonumber\\
&&\left.+\left[2\left(\frac{N-j+\frac{3}{2}}{2}\right)+2\left(\frac{N+j+\frac{3}{2}}{2}\right)\right]
\left(\frac{N+j+\frac{3}{2}}{2}\right)^{\frac{1}{2}}
\bd{b}^{\dagger}_{\frac{1}{2}(N-1,j-\frac{1}{2},\frac{1}{2})}
\bd{b}^{-\frac{1}{2}(N,j+\frac{1}{2},\frac{1}{2})}
\right\}\nonumber\\
&=&\frac{\gamma}{2}[2N+3]\widetilde{\bd{K}}_{+} \eeqa
\end{widetext}

Defining

\beqa
{\bd K}_\pm & = & \eta {\widetilde {\bd K}}_\pm
\nonumber \\
{\bd K}_0 & = & \eta^{2} {\widetilde {\bd K}}_0,
\eeqa
we arrive at the relation

\beqa
\left[ {\bd K}_0 , {\bd K}_+ \right] & = & \frac{\gamma}{2}
\left( 2N+3\right) \eta^{2} {\bd K}_+.
\eeqa
Choosing

\beqa
\eta & = & \left( \sqrt{ \frac{\gamma}{2} \left( 2N+3 \right) }
\right)^{-1}
\eeqa
we arrive at the well known $SU(2)$ algebra. Of course, this
has to be repeated for ${\bd K}_-$, which is easily obtained by
taking the
adjoint of ${\bd K}_+$.


\begin{thebibliography}{99}

\bibitem{adam1}
A. S. Szczepaniak and E. S. Swanson, Phys. Rev. D {\bf 65} (2001), 025012.

\bibitem{adam-peter} P. O. Hess, A. Szczepaniak, Phys. Rev. {\bf C73} (2006), 025201.

\bibitem{rmf} T. Y\'epez Mart\ii nez, P. O. Hess, A. Szczepaniak and
O. Civitarese, Rev. Mex. F\ii s. (2009), in press.

\bibitem{jutta}
Escher and J. P. Draayer, J. Math. Phys. {\bf 39}, 5123
(1998).

\bibitem{lee} T. D. Lee, {\it Particle Physics and Introduction
to Field Theory}, (World Scientific, Singapore, 1981).

\bibitem{draayer1}
J. P. Draayer and Y. Akiyama, J. Math. Phys. {\bf 14}, 1904
(1973).

\bibitem{draayer2}
Y. Akiyama and J. P. Draayer, Comp. Phys. Comm.
{\bf 5}, 405 (1973).

\bibitem{ring} P. Ring, P. Schuck, {\it The Nuclear Many Body Problem},
(Springer, Heidelberg, 1980).

\bibitem{double} S. Cohen and C. Tomase, {\it Systems of Bilinear Equations},
Report of  Computer Science Department, Standford University, 1995.

\bibitem{grand} T. DeGrand and C. DeTar, {\it Lattice Methods for Quantum
Chromodynamics}, (World Scientific, Singapore, 2006).

\bibitem{bahri1}
D. J. Rowe and C. Bahri, J. Math. Phys. {\bf 41}, 6544
(2000).

\bibitem{bahri2}
C. Bahri, D. J. Rowe and J. P. Draayer,
Comp. Phys. Comm. {\bf 159}, 121 (2004).


\end{thebibliography}
\end{document}